\documentstyle[12pt]{article}

\renewcommand{\baselinestretch}{1.0}
\newcommand{\sect}[1]{\setcounter{equation}{0}\section{#1}}

\newcommand{\app}[1]{\setcounter{section}{0}
\setcounter{equation}{0} \renewcommand{\thesection}{\Alph{section}}
\section{#1}}
\newcommand{\eq}{\begin{equation}}
\newcommand{\eqa}{\begin{eqnarray}}
\newcommand{\en}{\end{equation}}
\newcommand{\ena}{\end{eqnarray}}
\newcommand{\enn}{\nonumber \end{equation}}

\def\sk{\vskip .4cm}
\def\noi{\noindent}
\def\om{\omega}
\def\al{\alpha}
\def\be{\beta}
\def\ga{\gamma}
\def\Ga{\Gamma}
\def\de{\delta}
\def\del{\delta}
\def\la{\lambda}
\def\lam{{1 \over \lambda}}
\def\alb{\bar{\alpha}}
\def\beb{\bar{\beta}}
\def\gab{\bar{\gamma}}
\def\deb{\bar{\delta}}

\def\rhop{{\rho}^{\prime}}
\def\taup{{\tau}^{\prime}}
\def\rhopp{\rho ''}
\def\thetap{{\theta}^{\prime}}

\def\unmezzo{{1 \over 2}}
\def\epsi{\varepsilon}
\def\we{\wedge}

\def\de{\delta}

\def\part{\partial}

\def\pdxi{{\partial \over {\partial x^i}}}
\def\pdy#1{{\partial \over {\partial y^{#1}}}}
\def\pdx#1{{\partial \over {\partial x^{#1}}}}
\def\pdyx#1{{\partial \over {\partial (yx)^{#1}}}}

\def\R#1#2{ R^{#1}_{~~~#2} }
\def\Rp#1#2{ (R^+)^{#1}_{~~~#2} }
\def\Rm#1#2{ (R^-)^{#1}_{~~~#2} }
\def\Rinv#1#2{ (R^{-1})^{#1}_{~~~#2} }
\def\Rpm#1#2{(R^{\pm})^{#1}_{~~~#2} }
\def\Rpminv#1#2{((R^{\pm})^{-1})^{#1}_{~~~#2} }
\def\RRpm{R^{\pm}}

\def\Rhat#1#2{ \Rh^{#1}_{~~~#2} }
\def\Rhats#1#2{ \Rhs^{#1}_{~~~#2} }
\def\Rhatinv#1#2{ (\Rh^{-1})^{#1}_{~~~#2} }
\def\Z#1#2{ Z^{#1}_{~~~#2} }

\def\Rhs{{\hat R}}
\def\Rh{{\Lambda}}

\def\ff#1#2#3{f_{#1~~~#3}^{~#2}}
\def\MM#1#2#3{M^{#1~~~#3}_{~#2}}
\def\cchi#1#2{\chi^{#1}_{~#2}}
\def\ome#1#2{\om_{#1}^{~#2}}
\def\RRhat#1#2#3#4#5#6#7#8{\Rh^{~#2~#4}_{#1~#3}|^{#5~#7}_{~#6~#8}}
\def\RRhatinv#1#2#3#4#5#6#7#8{(\Rh^{-1})^
{~#2~#4}_{#1~#3}|^{#5~#7}_{~#6~#8}}
\def\U#1#2#3#4#5#6#7#8{U^{~#2~#4}_{#1~#3}|^{#5~#7}_{~#6~#8}}
\def\Cb{{\bf C}}
\def\CC#1#2#3#4#5#6{\Cb_{~#2~#4}^{#1~#3}|_{#5}^{~#6}}
\def\cc#1#2#3#4#5#6{C_{~#2~#4}^{#1~#3}|_{#5}^{~#6}}

\def\C#1#2{ {\bf C}_{#1}^{~~~#2} }
\def\c#1#2{ C_{#1}^{~~~#2} }

\def\Dmat#1#2{D^{#1}_{~#2}}
\def\Dmatinv#1#2{(D^{-1})^{#1}_{~#2}}
\def\DR{\Delta_R}
\def\DL{\Delta_L}
\def\f#1#2{ f^{#1}_{~~#2} }

\def\T#1#2{ T^{#1}_{~~#2} }
\def\Ti#1#2{ (T^{-1})^{#1}_{~~#2} }

\def\TP{ T^{\prime} }
\def\M#1#2{ M_{#1}^{~#2} }
\def\qm{q^{-1}}

\def\D{\Delta}

\def\Ap{A^{\prime}}
\def\Dp{\Delta^{\prime}}
\def\Ip{I^{\prime}}
\def\ep{\epsi^{\prime}}
\def\kp{\kappa^{\prime}}
\def\kpm{\kappa^{\prime -1}}
\def\km{\kappa^{-1}}
\def\gp{g^{\prime}}
\def\qone{$q \rightarrow 1~$}

\def\suq{$SU_q(2)~$}

\def\LL{L^*}
\def\ll#1{L^*_{#1}}
\def\RR{R^*}
\def\rr#1{R^*_{#1}}

\def\Lpm#1#2{(L^{\pm})^{#1}_{~~#2}}
\def\LLpm{L^{\pm}}
\def\LLp{L^{+}}
\def\LLm{L^{-}}
\def\Lp#1#2{(L^{+})^{#1}_{~~#2}}
\def\Lm#1#2{(L^{-})^{#1}_{~~#2}}

\def\Fun{$Fun(G)~$}
\def\invG{{}_{{\rm inv}}\Ga}
\def\Ginv{\Ga_{{\rm inv}}}
\def\qonelim{\stackrel{q \rightarrow 1}{\longrightarrow}}
\def\viel#1#2{e^{#1}_{~~{#2}}}

\begin{document}
\begin{titlepage}
\rightline{CERN--TH.6565/92}
\rightline{DFTT-22/92}
\vskip 2em
\begin{center}{\bf AN INTRODUCTION
TO NON-COMMUTATIVE}\\{\bf DIFFERENTIAL GEOMETRY ON QUANTUM GROUPS}\\[6em]
 Paolo Aschieri${}^{\diamond *}$
and Leonardo Castellani${}^{*}$ \\[2em]
{\sl${}^{\diamond}$ CERN, CH-1211 Geneva 23, Switzerland.\\
{}$^{*}$Istituto Nazionale di
Fisica Nucleare, Sezione di Torino\\
and\\Dipartimento di Fisica Teorica\\
Via P. Giuria 1, 10125 Torino, Italy.}  \\[6em]
\end{center}
\begin{abstract}
We give a pedagogical introduction to the differential calculus
on quantum groups by stressing at all stages the connection
with the classical case ($q \rightarrow 1$ limit). The Lie
derivative and the contraction operator on forms and tensor fields
are found. A new, explicit form of the Cartan--Maurer equations is
presented. The example of a bicovariant differential calculus on
the quantum group $GL_q(2)$ is given in detail. The softening of
a quantum group is considered, and we introduce $q$-curvatures satisfying
q-Bianchi identities, a basic ingredient for the construction of
$q$-gravity and $q$-gauge theories.
\end{abstract}

\vskip 3cm

\noi CERN--TH.6565/92

\noi DFTT-22/92

\noi July 1992

\noi \hrule
\vskip.2cm
\hbox{\vbox{\hbox{{\small{\it email addresses:}}}\hbox{}}
 \vbox{\hbox{{\small Decnet=(39163::ASCHIERI, CASTELLANI;
VXCERN::ASCHIERI)}}
\hbox{{\small Bitnet=(ASCHIERI, CASTELLANI@TORINO.INFN.IT )}}}}

\end{titlepage}
\newpage
\setcounter{page}{1}

\sect{Introduction}

Quantum groups \cite{Drinfeld}--\cite{Majid1} have emerged
%\cite{Drinfeld,Jimbo,FRT,Majid1} have emerged
as interesting non-trivial
generalizations of Lie groups. The latter are recovered in the limit $q
\rightarrow 1$, where $q$ is a continuous deformation parameter (or
set of parameters). In the $q \not= 1$ case, the $q$-group may have, and
in general does have, more than one corresponding $q$-algebra, the
$q$-analogue of the Lie algebra. This can be rephrased by saying
that the
differential calculus on $q$-groups is not unique \cite{Wor}. On a
classical Lie group one can define a left and right action of the group
on itself, and these commute. By imposing this ``bicovariance"
to hold also in the $q$-deformation, one restricts the possible
differential calculi (still to a number $> 1$ in general).
\sk
Our motivations for studying the differential geometry of quantum groups
are twofold:

-- the $q$-differential calculus offers a natural scenario for
$q$-deformations of gravity theories, based on the quantum Poincar\'e
group \cite{Cas1}. Space-time becomes
non-commutative, a fact that does not
contradict (Gedanken) experiments under the Planck length, and that
could possibly provide a regularization mechanism \cite{Connes,Majid}.

-- the quantum Cartan-Maurer
equations define $q$-curvatures, and these can
be used for constructing $q$-gauge theories \cite {Cas2}. Here
space-time can be taken
to be the ordinary commutative Minkowski spacetime, while the
$q$-structure resides on the fibre, the gauge potentials being
non-commuting. These theories could offer interesting examples of a
novel way of breaking symmetry by $q$-deforming the classical one.
\sk

In this paper we intend to give an introductory review of the
$q$-differential calculus. A discussion on Hopf structures will {\sl not}
be omitted: in Section 2 we recognize these structures in ordinary Lie
groups and Lie algebras. Quantum groups and their non-commutative
geometry are discussed in Sections 3 and 4, and Section 5 describes an
explicit construction of a bicovariant calculus on quantum groups.
Some new results of Section 5 include
a formula that gives
the commutations of the left-invariant one-forms, and (as a consequence)
a new expression for the Cartan-Maurer equations.

The example of $GL_q(2)$ [and its restrictions to $U_q(2)$ and \suq ]
is systematically used to illustrate
the general concepts. In Section 6 we study the
quantum Lie derivative, an essential tool for the definition of
$q$-variations. Section 7 extends the notion of ``soft" group manifolds
(see for example \cite{CDF}) to
$q$-groups, by introducing $q$-curvatures.
\sk
After refs. \cite{Wor}, there
have been a number of papers treating the differential calculus on
$q$-groups from various points of view
\cite
{Manin}--\cite{Sun}.

In this paper we use the
formalism of refs. \cite{Wor} and \cite{Jurco}.

\sect{Hopf structures in ordinary Lie groups and Lie algebras}

Let us begin by considering \Fun, the set of differentiable functions
from a Lie group $G$ into the complex numbers $\Cb$. \Fun  is an
algebra with the usual pointwise sum and
product $(f+h)(g)=f(g)+h(g),~(f \cdot h)=f(g) h(g),
{}~(\lambda f)(g)=\lambda f(g)$, for $f,h \in Fun(G),~g \in G,~\lambda \in
\Cb$. The unit of this algebra is $I$, defined by $I(g)=1,~\forall g \in
G$.

Using the group structure of $g$, we can introduce on \Fun three other
linear mappings, the coproduct $\D$, the counit $\epsi$, and
the coinverse
(or antipode) $\kappa$:
\begin{eqnarray}
\D (f)(g,\gp) &\equiv& f(g\gp),~~~\D:Fun(G) \rightarrow Fun(G)\otimes
  Fun(G) \label{cop} \\
\epsi(f) &\equiv& f(e),~~~~~~\epsi:Fun(G) \rightarrow \Cb \label{cou}\\
(\kappa f)(g) &\equiv& f(g^{-1}),~~~\kappa:Fun(G) \rightarrow Fun(G)
  \label{coi}
\end{eqnarray}

\noi where $e$ is the unit of $G$. It
Xis not difficult to verify the following properties of the co-structures:
\begin{eqnarray}
 & & (id \otimes \D)\D=(\D\otimes id)\D~~~({\rm coassociativity~of~\D})
\label{prop1}\\
 & & (id\otimes \epsi)\D(a)=(\epsi\otimes id)\D(a)=a \label{prop2}\\
 & & m(\kappa\otimes id)\D(a)=m(id\otimes\kappa)\D(a)=\epsi(a) I
\label{prop3}
\end{eqnarray}

\noi and
\begin{eqnarray}
 & & \D(ab)=\D(a)\D(b),~~~\D(I)=I\otimes I \label{prop4}\\
 & & \epsi(ab)~=\epsi(a) \epsi(b),~~~~~~\epsi(I)=1 \label{prop5} \\
 & & \kappa(ab) \; =\kappa(b)\kappa(a),~~~~~\kappa(I)=I \label{prop6}
\end{eqnarray}

\noi where $a,b \in A=Fun(G)$ and $m$ is the multiplication map
$m(a\otimes b)\equiv ab$. The product in $\D(a)\D(b)$ is the product in
$A\otimes A$: $(a\otimes b)(c\otimes d)=ab\otimes cd$.

In general a coproduct can be expanded on
$A \otimes A$ as:
\eq
\D(a)=\sum_i a_1^i \otimes a_2^i \equiv a_1 \otimes a_2, \label{not1}
\en
\noi where $a_1^i, a_2^i \in A$ and $a_1 \otimes a_2$ is a
short-hand notation we will often use in the sequel. For example for
$A=Fun(G)$ we have:
\eq
\D(f)(g,\gp)=(f_1 \otimes f_2)(g,\gp)=f_1(g)f_2(\gp)=f(g\gp).
    \label{not2}
\en
\noi Using (\ref{not2}), the proof of (\ref{prop1})-(\ref{prop3}) is
immediate.
\sk
An algebra $A$ endowed with the homomorphisms $\D:A \rightarrow A
\otimes A$ and $\epsi: A \rightarrow \Cb$, and the antimorphism
$\kappa: A\rightarrow A$ satisfying the properties
(\ref{prop1})-(\ref{prop6})
is a {\sl Hopf algebra}. Thus \Fun is a Hopf algebra.\footnote
{To be precise, \Fun is a Hopf algebra when $Fun(G \times G)$ can be
identified with $Fun(G) \otimes Fun(G)$, since only then can one define
a coproduct as in (\ref{cop}). This is possible for compact
$G$.} Note that the properties (\ref{prop1})-(\ref{prop6})
imply the relations:
\eqa
& & \D(\kappa (a))=\kappa (a_2) \otimes \kappa (a_1) \label{Dka}\\
& & \epsi (\kappa (a))=\epsi (a). \label{epsika}
\ena
\sk
Consider now the algebra $A$ of polynomials in the matrix
elements $\T{a}{b}$ of the fundamental representation of $G$. The
algebra $A$ is said
to be freely generated by the $\T{a}{b}$.

It is clear that
$A \subset Fun(G)$, since $\T{a}{b} (g)$ are
functions on $G$. In fact every function on $G$
can be expressed as a polynomial in the $\T{a}{b}$ (the reason
is that the matrix
elements of all irreducible representations of $G$ form a complete
basis of $Fun(G)$, and these matrix elements can be constructed out of
appropriate products of $\T{a}{b} (g)$), so that $A=Fun(G)$.
The group manifold $G$ can be completely characterized by $Fun(G)$, the
co-structures on \Fun carrying the information about the group
structure of $G$. Thus a classical Lie group can be ``defined" as
the algebra $A$ freely generated by the (commuting) matrix
elements $\T{a}{b}$ of
the fundamental representation of $G$, seen as functions on $G$.
This definition admits non-commutative generalizations, i.e. the
quantum groups discussed in the next Section.
\sk
Using the elements $\T{a}{b}$ we can write an
explicit formula for the expansion (\ref{not1}) or (\ref{not2}): indeed
(\ref{cop}) becomes
\eq
\D(\T{a}{b})(g,\gp)=\T{a}{b} (g\gp)=\T{a}{c}(g) \T{c}{b}(\gp),
\en
\noi since $T$ is a matrix representation of $G$. Therefore:
\eq
\D(\T{a}{b})=\T{a}{c} \otimes \T{c}{b}. \label{copT}
\en
Moreover, using (\ref{cou}) and (\ref{coi}), one finds:
\begin{eqnarray}
 & & \epsi(\T{a}{b})=\de^a_b \label{couT}\\
 & & \kappa(\T{a}{b})=\Ti{a}{b}. \label{coiT}
\end{eqnarray}
Thus the algebra $A=Fun(G)$ of polynomials in the elements $\T{a}{b}$ is
a Hopf algebra with co-structures defined by
(\ref{copT})-(\ref{coiT}) and (\ref{prop4})-(\ref{prop6}).
\sk
Another example of Hopf algebra is given by any ordinary Lie algebra,
or more precisely by the universal enveloping algebra of a Lie algebra,
i.e. the algebra (with unit $I$) of polynomials in the generators $T_i$
modulo the commutation relations
\eq
[T_i,T_j]=\c{ij}{k} T_k. \label{clcomm}
\en
Here we define the co-structures as:
\begin{eqnarray}
 & & \D(T_i)=T_i \otimes I + I \otimes T_i~~~\D(I)=I\otimes I
  \label{copL}\\
 & & \epsi(T_i)=0~~~~~~~~~~~~~~~~~~~~~~\epsi(I)=1 \label{couL}\\
 & & \kappa(T_i)=-T_i~~~~~~~~~~~~~~~~~~\kappa(I)=I \label{coiL}
\end{eqnarray}
\noi The reader can check that (\ref{prop1})-(\ref{prop3}) are
satisfied.
\sk
In general the dual of a (finite-dimensional) Hopf algebra $A$
is a Hopf algebra $\Ap$, whose structures and co-structures are given,
respectively, by the co-structures and structures of $A$, i.e.:
\begin{eqnarray}
& &  \chi_1 \chi_2 (a)\equiv (\chi_1 \otimes \chi_2)\D(a),
{}~~~\chi_1,\chi_2 \in \Ap \label{copd}\\
& & \Ip(a) \equiv \epsi(a)~~~~~~~~~~~~~~~~~~~~~\Ip={\rm unit~of~}\Ap
\label{coud}
\end{eqnarray}
\noi and:
\begin{eqnarray}
& & \Dp(\chi)(a\otimes b)\equiv \chi(ab) \label{copdual}\\
& & \ep(\chi) \equiv \chi(I) \label{coudual}\\
& & \kp(\chi)(a)\equiv \chi(\kappa(a)) \label{coidual}
\end{eqnarray}

\sect{Quantum groups. The example of $GL_q(2)$}

Quantum groups are introduced as non-commutative deformations of the
algebra $A=Fun(G)$ of the previous section [more precisely as
non-commutative Hopf algebras obtained by continuous deformations of
the Hopf algebra $A=Fun(G)$]. In the following we consider quantum
groups defined as the associative algebras $A$ freely
generated by non-commuting matrix entries
$\T{a}{b}$ satisfying the relation
\eq
\R{ab}{ef} \T{e}{c} \T{f}{d} = \T{b}{f} \T{a}{e} \R{ef}{cd} \label{RTT}
\en
\noi and some other conditions depending on which classical group
%(deformations of the classical $A_{n-1}$, $B_n$, $C_n$ and $D_n$
%Lie groups)
we are deforming (see later). The matrix $R$ controls the
non-commutativity of the $\T{a}{b}$, and its elements depend
continuously on a (in general complex) parameter $q$, or even a set of
parameters. For $q\rightarrow 1$, the so-called ``classical limit", we
have
\eq
\R{ab}{cd} \qonelim \de^a_c \de^b_d,  \label{limR}
\en
\noi i.e. the matrix entries $\T{a}{b}$ commute for $q=1$, and one
recovers the ordinary $Fun(G)$.
%In Appendix A we recall the
%explicit form of the $R$ matrix for the
%$A_{n-1}$, $B_n$, $C_n$ and $D_n$ $q$-groups, and some of its properties
%\cite{FRT}.

The associativity of $A$ implies a consistency condition on the $R$
matrix, the quantum Yang--Baxter equation:
\eq
\R{a_1b_1}{a_2b_2} \R{a_2c_1}{a_3c_2} \R{b_2c_2}{b_3c_3}=
\R{b_1c_1}{b_2c_2} \R{a_1c_2}{a_2c_3} \R{a_2b_2}{a_3b_3}. \label{YB}
\en
For simplicity we rewrite the ``RTT" equation (\ref{RTT}) and the
quantum Yang--Baxter equation as
\eq
R_{12} T_1 T_2 = T_2 T_1 R_{12} \label{rtt}
\en
\eq
R_{12} R_{13} R_{23}=R_{23} R_{13} R_{12}, \label{yb}
\en
\noi where the subscripts 1, 2 and 3 refer to different couples of
indices. Thus $T_1$ indicates the matrix $\T{a}{b}$, $T_1 T_1$
indicates $\T{a}{c} \T{c}{b}$, $R_{12} T_2$ indicates $\R{ab}{cd}
\T{d}{e}$ and so on, repeated subscripts meaning matrix
multiplication. The quantum Yang--Baxter equation (\ref{yb})
is a condition sufficient
for the consistency of the RTT equation (\ref{rtt}). Indeed
the product of three distinct elements $\T{a}{b}$, $\T{c}{d}$
and $\T{e}{f}$, indicated by $T_1T_2T_3$, can be reordered as
$T_3T_2T_1$ via two differents paths:
\eq
 T_1T_2T_3 \begin{array}{c} \nearrow\\ \searrow \end{array}
           \begin{array}{c} T_1T_3T_2 \rightarrow T_3T_1T_2 \\ {} \\
                            T_2T_1T_3 \rightarrow T_2T_3T_1 \end{array}
           \begin{array}{c} \searrow\\ \nearrow \end{array}
 T_3T_2T_1
\en
\noi by repeated use of the RTT equation. The relation (\ref{yb})
ensures that the two paths lead to the same result.
\sk
The algebra $A$ (``the quantum group") is a non-commutative Hopf algebra
whose co-structures are the same of those defined for the commutative
Hopf algebra \Fun of the previous section, eqs.
(\ref{copT})-(\ref{coiT}), (\ref{prop4})-(\ref{prop6}).
\sk
Let us give the example of $SL_q(2)$, the algebra freely generated by
the elements $\al,\be,\ga$ and $\de$ of the $2 \times 2$ matrix
\eq
\T{a}{b}= \left( \begin{array}{cc} \al & \be \\ \ga & \de \end{array}
    \right)   \label{Tmatrix}
\en
\noi satisfying the commutations
\eqa
\al\be=q\be\al,~~\al\ga=q\ga\al,~~\be\de=q\de\be,~~\ga\de=q\de\ga
\nonumber \\
\be\ga=\ga\be,~~\al\de-\de\al=(q-\qm)\be\ga,~~~~q\in\Cb
\label{sucomm}
\ena
\noi and
\eq
{\det}_q T\equiv\al\de-q\be\ga=I. \label{sudet}
\en
The commutations (\ref{sucomm}) can be obtained from (\ref{RTT}) via the
$R$ matrix
\eq
\R{ab}{cd}=\left( \begin{array}{cccc} q & 0 & 0 & 0 \\
                                      0 & 1 & 0 & 0 \\
                                      0 & q-\qm & 1 & 0 \\
                                      0 & 0 & 0 & q  \end{array} \right)
  \label{Rsu}
\en
\noi where the rows and columns are numbered in the order
11, 12, 21, 22.

It is easy to verify that the ``quantum determinant" defined in
(\ref{sudet}) commutes with $\al,\be,\ga$ and $\de$, so that the
requirement $\det_q T=I$ is consistent. The matrix inverse of $\T{a}{b}$
is
\eq
\Ti{a}{b}= ({\det}_q T)^{-1} \left( \begin{array}{cc} \de & -\qm\be \\
          -q\ga & \al
         \end{array} \right)   \label{Timatrix}
\en
\sk
The coproduct, counit and coinverse of $\al,\be,\ga$ and $\de$ are
determined via formulas (\ref{copT})-(\ref{coiT}) to be:
\eqa
\D(\al)=\al\otimes\al+\be\otimes\ga,~~~\D(\be)=\al\otimes\be+
\be\otimes\de \nonumber\\
\D(\ga)=\ga\otimes\al+\de\otimes\ga,~~~\D(\de)=\ga\otimes\be+
\de\otimes\de \label{copsu}\\
\epsi(\al)=\epsi(\de)=1,~~~\epsi(\be)=\epsi(\ga)=0 \label{cousu}\\
\kappa(\al)=\de,~~\kappa(\be)=\qm \be,~~\kappa(\ga)=-q\ga,~~
\kappa(\de)=\al . \label{coisu}
\ena
\sk

{\sl Note 1:} In general $\kappa^2 \not= 1$, as can be seen from
(\ref{coisu}). The following useful relation holds \cite{FRT}:
\eq
\kappa^2 (\T{a}{b})=\Dmat{a}{c} \T{c}{d} \Dmatinv{d}{b}=d^a d^{-1}_b
\T{a}{b},  \label{k2}
\en
\noi where $D$ is a diagonal matrix, $\Dmat{a}{b}=d^a \de^a_{b}$, given
by $d^a=q^{2a-1}$ for the $q$-groups $A_{n-1}$.
\sk
{\sl Note 2:} The commutations (\ref{sucomm}) are compatible with the
coproduct $\D$, in the sense that $\D(\al\be)=q\D(\be\al)$ and so on.
In general we must have
\eq
\D(R_{12}T_1T_2)=\D(T_2T_1R_{12}), \label{DRTT}
\en
\noi which is easily verified using $\D(R_{12}T_1T_2)=R_{12}\D(T_1)
\D(T_2)$ and $\D(T_1)=T_1 \otimes T_1$. This is equivalent to proving
that the matrix elements of the matrix product $T_1\TP_1$, where
$\TP$ is a matrix [satisfying (\ref{RTT})]
whose elements {\sl commute} with those of $\T{a}{b}$,
still obey the commutations (\ref{rtt}).
\sk
{\sl Note 3:} $\D(\det_qT)=\det_qT\otimes \det_qT~$ so that the coproduct
property $\D(I)=I\otimes I$ is compatible with $\det_qT=I$.
\sk
{\sl Note 4:} Other conditions compatible with the RTT relation can be
imposed on $\T{a}{b}$:

i) Unitarity condition: $T^\dagger=T^{-1}
\Rightarrow \alb=\de,~\beb=-q\ga
,~\gab=-\qm\be,~\deb=\al$, where $q$ is a real number and
the bar denotes an involution, the $q$-analogue of complex
conjugation, satisfying $\overline{(\al\be)}=\beb\alb$ etc.
Restricts $SL_q(2)$ to $SU_q(2)$.

ii) Reality condition: ${\bar T}=T \Rightarrow \alb=\al,~\beb=\be,~\gab=
\ga,~\deb=\de$, $|q|=1$. Restricts $SL_q(2)$ to $SL_q(2,R)$.

iii) The $q$-analogue of orthogonal and symplectic groups can also be
defined, see \cite{FRT}.
\sk
{\sl Note 5:} The condition (\ref{sudet}) can be relaxed. Then we have
to include the central element $\zeta=(\det_q T)^{-1}$ in $A$, so
as to be able
to define the inverse of the $q$-matrix $\T{a}{b}$
as in (\ref{Timatrix}),
and the coinverse of the element $\T{a}{b}$ as in (\ref{coiT}). The
$q$-group is then $GL_q(2)$, and the unitarity condition restricts it to
$U_q(2)$. The reader can deduce the co-structures on $\zeta$: $\D(\zeta)
=\zeta \otimes \zeta,~\epsi(\zeta)=1,~\kappa(\zeta)=\det_q T$.
\sk
{\sl Note 6:} More generally, the quantum determinant of $n \times n$
$q$-matrices is defined by
$\det_qT=\sum_{\sigma} (-q)^{l(\sigma)} \T{1}{\sigma (1)} \cdots
\T{n}{\sigma (n)}$, where $l(\sigma)$ is the minimal number of inversions
in the permutation $\sigma$. Then $\det_q T=1$ restricts $GL_q(n)$ to
$SL_q(n)$.
\sk
{\sl Note 7:} We recall the important relations \cite{FRT}
for the $\Rhs$ matrix
defined by $\Rhats{ab}{cd} \equiv \R{ba}{cd}$, whose $q=1$ limit is the
permutation operator $\delta^a_d \delta^b_c$:
\eq
\Rhs^2=(q-q^{-1}) \Rhs + I,~~{\rm for}~A_{n-1} ~~~{\rm(Hecke~condition)}
  \label{Hecke}
\en
\eq
(\Rhs -qI)(\Rhs +q^{-1} I)(\Rhs -q^{1-N}I)=0,~~{\rm for}~B_n,C_n,D_n,
  \label{R3}
\en
\noi with $N=2n+1$ for the series $B_n$ and $N=2n$ for $C_n$ and $D_n$.
 Moreover for all $A,B,C,D$ $q$-groups
the $R$ matrix is lower triangular
and satisfies:
\eq
\Rinv{ab}{cd} (q) = \R{ab}{cd} (q^{-1})
\en
\eq
\R{ab}{cd}=\R{dc}{ba}.
\en
\sect{Differential calculus on quantum groups}

In this section we give a short review of the bicovariant differential
calculus on $q$-groups as developed by Woronowicz \cite{Wor}. The
\qone limit will constantly appear in our discussion, so as to make
clear which classical structure is being $q$-generalized.
\sk
Consider the algebra $A$ of the preceding section, i.e. the algebra
freely generated by the matrix entries $\T{a}{b}$, modulo the relations
(\ref{RTT}) and possibly some reality or orthogonality conditions.
\sk
A {\bf {\sl first-order differential calculus}} on $A$ is then defined by
\sk
i) a linear map $d$: $A \rightarrow \Gamma$, satisfying the Leibniz rule
\eq
d(ab)=(da)b+a(db),~~\forall a,b\in A; \label{Leibniz}
\en
$\Gamma$ is an
appropriate bimodule (see for example \cite{Abe}) on $A$, which
essentially means that its elements can be
multiplied on the left and on the right by elements of $A$, and
$q$-generalizes the space of 1-forms on a Lie group;

ii) the possibility of expressing any $\rho \in \Ga$ as
\eq
\rho=a_k db_k \label{adb}
\en
\noi for some $a_k,b_k$ belonging to $A$.
\sk
\noi {\sl{\bf Left- and right-covariance}}
\sk
The first-order differential calculus $(\Ga,d)$ is said to be {\sl
left- and right-covariant} if we can consistently
define a left and right action of
the $q$-group on $\Ga$ as follows
\eqa
& & \DL(adb)=\D(a)(id \otimes d)\D(b),~~~\DL:\Ga\rightarrow A\otimes \Ga
{}~~~{\rm (left covariance)} \label{leftco}\\
& & \DR(adb)=\D(a)(d \otimes id)\D(b),~~~\DR:\Ga\rightarrow \Ga\otimes A
{}~~~{\rm (right covariance)} \label{rightco}
\ena

How can we understand these left and right actions on $\Ga$ in the \qone
limit ? The first observation is that the coproduct $\D$ on $A$ is
directly related, for $q=1$, to the pullback induced by left
multiplication of the group on itself
\eq
L_x y \equiv xy,~~~\forall x,y \in G. \label{leftmu}
\en
\noi This induces the left action (pullback)
$\ll{x}$ on the functions on $G$:
\eq
\ll{x} f(y)\equiv f(xy)|_y,~~~~\ll{x} : Fun(G) \rightarrow Fun(G)
\label{pullback}
\en
\noi where $f(xy)|_y$ means $f(xy)$ seen as a function of $y$. Let us
introduce the mapping $\LL$ defined by
\eqa
&(\LL f) (x,y) \equiv (\ll{x} f)(y) = f(xy)|_y&  \nonumber\\
&\LL : Fun(G) \rightarrow Fun(G \times G) \approx Fun(G) \otimes Fun(G).
&
\label{LL}
\ena
The coproduct $\D$ on $A$, when $q=1$, reduces to the mapping $\LL$.
Indeed, considering $\T{a}{b}(y)$ as a function on $G$, we have:
\eq
\LL (\T{a}{b}) (x,y) = \ll{x} \T{a}{b}(y) = \T{a}{b} (xy)=\T{a}{c}(x)
\T{c}{b}(y),
\en
\noi since $\T{a}{b}$ is a representation of $G$. Therefore
\eq
\LL (\T{a}{b}) = \T{a}{c} \otimes \T{c}{b}
\en
\noi and $\LL$ is seen to coincide with $\D$, cf. (\ref{copT}).
\sk
The pullback $\ll{x}$ can also be defined on 1-forms $\rho$ as
\eq
(\ll{x} \rho) (y) \equiv \rho(xy)|_y
\en
\noi and here too we can define $\LL$ as
\eq
(\LL \rho)(x,y) \equiv (\ll{x} \rho)(y) = \rho(xy)|_y .
\en
In the $q=1$ case we are now discussing, the left action $\DL$ coincides
with this mapping $\LL$ for 1-forms. Indeed for $q=1$
\eqa
\lefteqn{\DL (adb) (x,y)=[\D (a) (id \otimes d) \D (b)] (x,y)=
[(a_1 \otimes a_2)
(id\otimes d)(b_1 \otimes b_2)](x,y)} \nonumber\\
& & =[a_1b_1\otimes  a_2db_2](x,y)=a_1(x)b_1(x) a_2(y)db_2(y)=
a_1(x)a_2(y)
d_y[b_1(x)b_2(y)]\nonumber\\
& & =\LL(a)(x,y) d_y [\LL(b) (x,y)]=a(xy) db(xy)|_y .
\ena
\noi On the other hand:
\eq
\LL(adb)(x,y)=a(xy) db(xy)|_y,
\en
\noi so that $\DL \rightarrow \LL$ when $q \rightarrow 1$. In
the last equation
we have
used the well-known property $\ll{x}(adb)=\ll{x}(a)\ll{x}(db)=
\ll{x}(a) d\ll{x}(b)$ of the
classical pullback. A similar discussion holds for $\DR$, and we have
$\DR \rightarrow \RR$ when \qone, where $\RR$ is defined via the
pullback $\rr{x}$ on
functions (0-forms) or on 1-forms induced by the right multiplication:
\eq
R_x y=yx,~~~\forall x,y \in G
\en
\eq
(\rr{x} \rho) (y)= \rho (yx)|_y
\en
\eq
(\RR \rho)(y,x) \equiv (\rr{x} \rho)(y).
\en
These observations explain why $\DL$ and $\DR$ are called left and right
actions of the quantum group on $\Ga$ when $q\not= 1$.
\sk
{}From the definitions (\ref{leftco}) and (\ref{rightco}) one deduces
the following properties \cite{Wor}:
\eq
(\epsi \otimes id) \DL (\rho)=\rho,~~~(id \otimes \epsi) \DR (\rho)=
\rho \label{Dprop1}
\en
\eq
(\D \otimes id)\DL=(id\otimes\DL)\DL,~~~(id\otimes\D)\DR=(\DR\otimes
 id)\DR \label{Dprop2}
\en
\sk
\noi {\sl {\bf Bicovariance}}
\sk
The left- and right-covariant calculus is said to be {\sl
bicovariant} when
\eq
(id \otimes \Delta_R) \Delta_L = (\Delta_L \otimes id) \Delta_R,
\label{bicovariance}
\en
which is the $q$-analogue of the fact that
left and right actions commute for $q=1$
($L^*_x R^*_y = R^*_y L^*_x$).
\sk\sk
\noi{\sl{\bf Left- and right-invariant $\om$}}
\sk
An element $\om$ of
$\Ga$ is said to be {\sl left-invariant} if
\eq
\DL (\om) = I \otimes \om \label{linvom}
\en
\noi and {\sl right-invariant} if
\eq
\DR (\om) = \om \otimes I. \label{rinvom}
\en
This terminology is easily understood: in the classical limit,
\eqa
\LL \om = I \otimes \om \\
\RR \om = \om \otimes I
\ena
\noi indeed define respectively left- and right-invariant 1-forms.

{\sl Proof:} the classical definition of left-invariance is
\eq
(\ll{x} \om)(y) = \om(y)
\en
\noi or, in terms of $\LL$,
\eq
(\LL\om)(x,y) = \ll{x}\om(y)=\om(y).
\en
\noi But
\eq
(I\otimes \om)(x,y)=I(x) \om(y)=\om(y),
\en
\noi so that
\eq
\LL \om = I \otimes \om
\en
\noi for left-invariant $\om$. A similar argument holds for
right-invariant $\om$.
\sk\sk
\noi {\sl{\bf Consequences}}
\sk
For any bicovariant first-order calculus one can prove the following
\cite{Wor}:
\sk
i) Any $\rho \in \Ga$ can be uniquely written in the form:
\eqa
\rho=a_i \om^i \label{rhoaom}\\
\rho=\om^i b_i \label{rhoomb}
\ena
\noi with $a_i$, $b_i \in A$, and {$\om^i$} a basis of
$\invG$, the linear
subspace of all left-invariant elements of $\Ga$. Thus, as in the
classical case, the whole of $\Ga$ is generated by a basis of left
invariant $\om^i$. An analogous theorem holds with a basis of right
invariant elements
$\eta^i \in \Ginv$. Note that in the quantum case we have
$a \om^i \not= \om^i a$ in general, the bimodule structure of
$\Ga$ being non-trivial for $q \not= 1$.
\sk
ii) There exist linear functionals $\f{i}{j}$ on $A$ such that
\eqa
& & \om^i b= (\f{i}{j} * b) \om^j \equiv (id \otimes \f{i}{j})
\Delta (b)\om^j \label{omb}\\
& & a\om^i=\om^j [(\f{i}{j} \circ \kappa^{-1})* a] \label{aom}
\ena
\noi for any $a,b \in A$. In particular,
\eq
\om^i \T{a}{b}=(id \otimes \f{i}{j})(\T{a}{c} \otimes \T{c}{b})
\om^j=\T{a}{c} \f{i}{j} (\T{c}{b}) \om^j. \label{omT}
\en
\noi Once we have the functionals $\f{i}{j}$, we
know how to commute elements of $A$ through elements of $\Ga$. The
$\f{i}{j}$ are uniquely determined by (\ref{omb}) and for consistency
must satisfy the conditions:
\eqa
& & \f{i}{j} (ab)= \f{i}{k} (a) \f{k}{j} (b) \label{propf1}\\
& & \f{i}{j} (I) = \del^i_j \label{propf2}\\
& & (\f{k}{j} \circ \kappa ) \f{j}{i} = \del^k_i ~\epsi;~~~
   \f{k}{j} (\f{j}{i} \circ \kappa)  = \del^k_i ~\epsi,~~~
   \label{propf3}
\ena
\noi so that their coproduct, counit and coinverse are given by:
\eqa
& & \Dp (\f{i}{j})=\f{i}{k} \otimes \f{k}{j}   \label{copf}\\
& & \ep (\f{i}{j}) = \del^i_j  \label{couf}\\
& & \kp (\f{k}{j}) \f{j}{i}= \de^k_i \epsi = \f{k}{j} \kp (\f{j}{i})
\label{coif}
\ena
\noi cf. (\ref{copdual})-(\ref{coidual}). Note
that in the $q=1$ limit $\f{i}{j}
\rightarrow \de^i_j \epsi$, i.e. $\f{i}{j}$ becomes proportional to the
identity functional $\epsi(a)=a(e)$, and formulas (\ref{omb}),
(\ref{aom}) become trivial, e.g. $\om^i b = b\om^i$ [use $\epsi * a=a$
from (\ref{prop2})].
\sk
iii) There exists an {\sl adjoint representation} $\M{j}{i}$ of the
quantum group, defined by the right action on the (left-invariant)
$\om^i$:
\eq
\DR (\om^i) = \om^j \otimes \M{j}{i},~~~\M{j}{i} \in A. \label{adjoint}
\en
It is easy to show that $\DR (\om^i)$ belongs to
$\invG \otimes A$, which proves the existence of $\M{j}{i}$.
In the classical case, $\M{j}{i}$ is indeed the adjoint representation
of the group. We
recall that in this limit the left-invariant 1-form $\om^i$ can be
constructed as
\eq
\om^i(y)T_i = (y^{-1} dy)^i T_i ,~~~~~~y\in G.
\en
Under right multiplication by a (constant) element $x \in G: y
\rightarrow yx$ we have, \footnote{Recall the $q=1$ definition of the
adjoint
representation $x^{-1} T_j x \equiv \M{j}{i} (x)T_i$.}
\eqa
&\om^i(yx)T_i=[x^{-1} y^{-1} d(yx)]^i T_i=[x^{-1}(y^{-1}dy)x]^iT_i\\
&=[x^{-1}T_j x]^i(y^{-1}dy)^j T_i=\M{j}{i}(x)\om^j(y) T_i,
\ena
\noi so that
\eq
\om^i(yx)=\om^j(y) \M{j}{i}(x)
\en
\noi or
\eq
\RR \om^i(y,x)=\om^j \otimes \M{j}{i}(y,x),
\en
\noi which reproduces (\ref{adjoint}) for $q=1$.

The co-structures on the $\M{j}{i}$ can be deduced \cite{Wor}:
\eqa
& & \Delta (\M{j}{i}) = \M{j}{l} \otimes \M{l}{i} \label{copM}\\
& & \epsi (\M{j}{i}) = \delta^i_j \label{couM}\\
& & \kappa (\M{i}{l}) \M{l}{j}=\delta^j_i=\M{i}{l} \kappa (\M{l}{j}).
\label{coiM}\ena
\noi For example, in order to find
the coproduct (\ref{copM}) it is sufficient to
apply $(id \otimes \D)$ to both members of (\ref{adjoint}) and use
the second of eqs.(\ref{Dprop2}).

The elements $\M{j}{i}$ can be used to
build a right-invariant basis of $\Ga$. Indeed the $\eta^i$ defined by
\eq
\eta^i \equiv \om^j \kappa (\M{j}{i}) \label{eta}
\en
are a basis of $\Ga$ (every element of $\Ga$ can be uniquely written
as $\rho = \eta^i b_i$) and their right-invariance can be checked
directly:
\eqa
\lefteqn{
\DR (\eta^i)=\DR (\om^j) \D [\kappa (\M{j}{i})] =\nonumber}\\
& & [\om^k \otimes \M{k}{j} ] [\kappa (\M{s}{i}) \otimes \kappa (
\M{j}{s})]=\om^k \kappa (\M{s}{i} ) \otimes \de^s_k I = \eta^i \otimes I
\ena
\noi It can be shown that the functionals $\f{i}{j}$ previously defined
satisfy:
\eqa
& & \eta^i b = (b * \f{i}{j} \circ \kappa^{-2}) \eta^j\\
& & a \eta^i = \eta^j [a*(\f{i}{j} \circ \kappa^{-1})],
\ena
\noi where $a*f \equiv (f \otimes id) \D(a),~f\in \Ap$.

Moreover, from the last of these relations, using (\ref{eta})
and (\ref{aom})
one can prove the relation
\eq
\M{i}{j} (a * \f{i}{k})=(\f{j}{i} * a) \M{k}{i}, \label{propM}
\en
\noi with $a* \f{i}{j} \equiv (\f{i}{k} \otimes id) \D(a)$.
\sk
iv) An {\sl exterior product}, compatible with the left
and right actions of the $q$-group, can be defined by a bimodule
automorphism
$\Rh$ in $\Gamma \otimes \Gamma$ that generalizes the
ordinary permutation operator:
\eq
\Rh (\om^i \otimes \eta^j )= \eta^j \otimes \om^i, \label{Rhat}
\en
\noi where $\om^i$ and $\eta^j$ are respectively left and right
invariant elements of $\Ga$. Bimodule automorphism means that
\eq
\Rh(a\tau)=a\Rh (\tau) \label{bimauto1}
\en
\eq
\Rh(\tau b)=\Rh(\tau)b \label{bimauto2}
\en
\noi for any $\tau \in \Ga \otimes \Ga$ and $a,b \in A$.
The tensor product between elements $\rho,\rhop \in \Ga$
is defined to
have the properties $\rho a\otimes \rhop=\rho \otimes a \rhop$, $a(\rho
\otimes \rhop)=(a\rho) \otimes \rhop$ and $(\rho \otimes \rhop)a=\rho
\otimes (\rhop a)$. Left and right actions on $\Ga \otimes \Ga$ are
defined by:
\eq
\DL (\rho \otimes \rhop)\equiv \rho_1 \rhop_1 \otimes \rho_2 \otimes
\rhop_2,~~~\DL: \Ga \otimes \Ga \rightarrow A\otimes\Ga\otimes\Ga
\label{DLGaGa}
\en
\eq
\DR (\rho \otimes \rhop)\equiv \rho_1 \otimes \rhop_1 \otimes \rho_2
\rhop_2,~~~\DR: \Ga \otimes \Ga \rightarrow \Ga\otimes\Ga\otimes A
\label{DRGaGa}
\en
\noi where as usual $\rho_1$, $\rho_2$, etc., are defined by
\eq
\DL (\rho) = \rho_1 \otimes \rho_2,~~~\rho_1\in A,~\rho_2\in \Ga
\en
\eq
\DR (\rho) = \rho_1 \otimes \rho_2,~~~\rho_1\in \Ga,~\rho_2\in A.
\en
\noi More generally, we can define the action of $\DL$
on $\Ga \otimes \Ga \otimes \cdots \otimes \Ga$ as

\[
\DL (\rho \otimes \rhop \otimes \cdots \otimes \rhopp)\equiv
\rho_1 \rhop_1 \cdots \rhopp_1 \otimes \rho_2 \otimes
\rhop_2\otimes \cdots \otimes \rhopp_2
\]
\eq
\DL: \Ga \otimes \Ga \otimes \cdots\otimes \Ga \rightarrow
A\otimes\Ga\otimes\Ga\otimes\cdots\otimes\Ga
\label{DLGaGaGa}
\en
\[
\DR (\rho \otimes \rhop \otimes \cdots \otimes \rhopp)\equiv
\rho_1 \otimes \rhop_1 \cdots \otimes \rhopp_1 \otimes \rho_2
\rhop_2 \cdots \rhopp_2
\]
\eq
\DR: \Ga \otimes \Ga \otimes \cdots\otimes \Ga \rightarrow
\Ga\otimes\Ga\otimes\cdots\otimes\Ga\otimes A.
\label{DRGaGaGa}
\en

\noi Left-invariance on $\Ga\otimes\Ga$ is naturally defined as
$\DL (\rho \otimes \rhop) = I \otimes \rho \otimes \rhop$ (similar
definition for right-invariance), so that, for example, $\om^i \otimes
\om^j$ is left-invariant, and is in fact a left-invariant basis for $\Ga
\otimes \Ga$.

-- In general
$\Rh^2 \not= 1$, since $\Rh (\eta^j \otimes \om^i )$ is not
necessarily equal to $ \om^i \otimes \eta^j $. By linearity, $\Rh$ can
be extended to the whole of $\Gamma \otimes \Gamma$.

-- $\Rh$ is invertible and commutes with the left and right action of
$q$-group $G$, i.e. $\DL \Rh (\rho \otimes \rhop)=(id \otimes \Rh) \DL
(\rho \otimes \rhop)
= \rho_1\rhop_1 \otimes \Rh(\rho_2 \otimes \rhop_2)$, and
similar for $\DR$. Then we see that $\Rh (\om^i \otimes \om^j)$ is
left-invariant, and therefore can be expanded on the left-invariant
basis $\om^k \otimes \om^l$:
\eq
\Rh (\om^i \otimes \om^j)= \Rhat{ij}{kl} \om^k \otimes \om^l.
\en
\sk
-- From the definition (\ref{Rhat}) one can prove that \cite{Wor}:
\eq
\Rhat{ij}{kl} = \f{i}{l} (\M{k}{j}); \label{RfM}
\en
\noi thus the functionals $\f{i}{l}$ and the elements $\M{k}{j} \in A$
characterizing the bimodule $\Gamma$ are dual in
the sense of eq. (\ref{RfM})
and determine the exterior product:
\eqa
& & \rho \we \rho ' \equiv \rho \otimes \rho ' -
\Rh (\rho \otimes \rho ')\\
& & \om^i \we \om^j \equiv \om^i \otimes \om^j -
\Rhat{ij}{kl} \om^k \otimes \om^l. \label{exom}
\ena
\noi Notice that, given the tensor $\Rhat{ij}{kl}$, we can compute the
exterior product of any $\rho,\rhop \in \Ga$, since
any $\rho \in \Ga$ is
expressible in terms of $\om^i$ [cf. (\ref{rhoaom}), (\ref{rhoomb})].
The classical limit of $\Rhat{ij}{kl}$ is
\eq
\Rhat{ij}{kl} \qonelim \de^i_l \de^j_k \label{limRhat}
\en
\noi since $\f{i}{j} \qonelim \de^i_l \epsi$ and $\epsi(\M{j}{k})=\de^
j_k$. Thus in the $q=1$ limit the product defined in
(\ref{exom}) coincides
with the usual exterior product.

{}From the property (\ref{bimauto1})
applied to the case $\tau=\om^i \otimes \om^j$, one can derive the
relation
\eq
\Rhat{nm}{ij} \f{i}{p} \f{j}{q} = \f{n}{i} \f{m}{j} \Rhat{ij}{pq}.
\label{Rff}
\en
\noi Applying both members of this equation to the element $\M{r}{s}$
yields the quantum Yang--Baxter equation for $\Rh$:
\eq
\Rhat{nm}{ij} \Rhat{ik}{rp} \Rhat{js}{kq}=\Rhat{nk}{ri} \Rhat{ms}{kj}
\Rhat{ij}{pq}, \label{QYB}
\en
\noi which is sufficient for the consistency of (\ref{Rff}).

Taking $a=
\M{p}{q}$ in (\ref{propM}), and using (\ref{RfM}), we find the relation
\eq
\M{i}{j} \M{r}{q} \Rhat{ir}{pk} = \Rhat{jq}{ri} \M{p}{r} \M{k}{i}
\label{RMM}
\en
\noi and, defining
\eq
\R{ji}{kl} \equiv \Rhat{ij}{kl}, \label{defR}
\en
\noi we see that $\M{i}{j}$ satisfies a relation identical to the
``RTT" equation (\ref{RTT}) for $\T{a}{b}$, and that $\R{ij}{kl}$
satisfies the quantum Yang--Baxter equation (\ref{YB}), sufficient
for the
consistency of (\ref{RMM}).  The range of the indices is
different, since {\small {\it i, j, ...}} are adjoint indices whereas
{\small {\it a, b ...}} are
in the fundamental representation of $G_q$.

\sk
v) Having the exterior product we can define the {\sl exterior
differential}
\eq
d~:~\Gamma \rightarrow \Gamma \we \Gamma
\en
\eq
d (a_k db_k) = da_k \we db_k,
\en
\noi which can easily be extended to
$\Gamma^{\we n}$ ($d: \Gamma^{\we n}
\rightarrow \Gamma^{\we (n+1)}$, $\Gamma^{\we n}$ being
defined as in the
classical case but with the quantum permutation
operator $\Rh$ \cite{Wor}) and has the following properties:
\eq
d(\theta \we \thetap)=d\theta \we \thetap + (-1)^k \theta \we d\thetap
\label{propd1}
\en
\eq
d(d\theta)=0\label{propd2}
\en
\eq
\DL (d\theta)=(id\otimes d)\DL(\theta)\label{propd3}
\en
\eq
\DR (d\theta)=(d\otimes id)\DR(\theta),\label{propd4}
\en
\noi where $\theta \in \Ga^{\we k}$, $\thetap \in \Ga^{\we n}$.
The last two properties express the fact that $d$ commutes with the left
and right action of the quantum group, as in the classical case.
\sk
vi) The space dual to the left-invariant subspace $\invG$ can be
introduced as a linear subspace of $\Ap$, whose basis elements $\chi_i
\in \Ap$ are defined by
\eq
da=(\chi_i * a)\om^i,~~~\forall a\in A. \label{defchi}
\en
\noi In order to reproduce the classical limit
\eq
da=\pdy{\mu} a(y) dy^{\mu} = \left(\pdy{\mu}a\right)
\viel{\mu}{i}(y) \viel{i}{\nu}
(y) dy^{\nu}=\left(\pdy{\mu}a\right) \viel{\mu}{i} (y) \om^i(y),
 \label{daclass}
\en
\noi where $\viel{i}{\nu} (y)$ is the vielbein of the group manifold
(and $\viel{\mu}{i}$ is its inverse), we must require
\eq
\chi_i(a) \qonelim \pdx{i} a(x) |_{x=e}. \label{chiclass}
\en
\noi Indeed, for $q=1$ we have
\eqa
\lefteqn{(\chi_i * a)(y) =  (id \otimes \chi_i) \LL (a) (y) = (id \otimes
  \chi_i)(a_1 \otimes a_2)(y) =a_1(y)\chi_i(a_2)=}\nonumber\\
 & & a_1 (y) [\pdx{i} a_2(x)]|_{x=e} = \pdx{i} [a_1(y) a_2(x)]|_{x=e}=
   \pdx{i} [(\LL a)(y,x)]|_{x=e}=\nonumber\\
 & & \pdx{i} [a(yx)]|_{x=e}=\pdyx{\mu} a(yx)|_{x=e} \pdx{i} (yx)^{\mu}|_
{x=e}=(\pdy{\mu} a(y)) \viel{\mu}{i} (y)
\ena
\noi and we recover (\ref{daclass}). In other words
\eq
\chi_i * a \qonelim \pdy{\mu} a(y) \viel{\mu}{i} \equiv \part_i a(y),
\label{chiaclass}
\en
\noi so that $\chi_i *$ is the $q$-analogue of left-invariant vector
fields, while $\chi_i$ is the $q$-analogue of the tangent vector at the
origin $e$ of $G$.
\sk
vii) The $\chi_i$ functionals close on the ``quantum Lie algebra":
\eq
\chi_i \chi_j - \Rhat{kl}{ij} \chi_k \chi_l = \C{ij}{k} \chi_k,
\label{qLie}
\en
\noi with $\Rhat{kl}{ij}$ as given in (\ref{RfM}). The product
$\chi_i\chi_j$ is defined by
\eq
\chi_i\chi_j \equiv (\chi_i \otimes \chi_j) \D
\en
\noi and sometimes indicated by $\chi_i * \chi_j$. Note that this $*$
product (called also convolution product) is associative:
\eq
\chi_i * (\chi_j * \chi_k)= (\chi_i * \chi_j)*\chi_k
\en
\eq
\chi_i * (\chi_j * a) = (\chi_i * \chi_j) * a,~~a\in A.
\en
\noi We leave the easy proof to the reader. The
$q$-structure constants $\C{ij}{k}$ are given by
\eq
\C{ij}{k} = \chi_j(\M{i}{k}). \label{Cijk}
\en
\noi This last equation is easily seen to hold in the $q=1$ limit, since
the $(\chi_j)_i^{~k} \equiv \C{ij}{k}$ are indeed in this case
the infinitesimal generators of the adjoint representation:
\eq
\M{i}{k}=\de^k_i + \C{ij}{k} x^j + 0(x^2).
\en
\noi Using $\chi_j \qonelim \pdx{j} |_{x=e}$ indeed yields (\ref{Cijk}).

By applying both sides of (\ref{qLie}) to $\M{r}{s} \in A$, we find the
$q$-Jacobi identities:
\eq
\C{ri}{n} \C{nj}{s}-\Rhat{kl}{ij} \C{rk}{n} \C{nl}{s} =
\C{ij}{k} \C{rk}{s},
\label{Jacobi}
\en
\noi which give an explicit matrix realization (the adjoint
representation) of the generators $\chi_i$:
\eq
(\chi_i)_k^{~l}=\chi_i(\M{k}{l})=\C{ki}{l}. \label{chiadj}
\en
\noi Note that the $q$-Jacobi identities (\ref{Jacobi}) can also be given
in terms of the $q$-Lie algebra generators $\chi_i$ as :
\eq
[[\chi_r,\chi_i],\chi_j]-\Rhat{kl}{ij} [[\chi_r,\chi_k],\chi_l]=
[\chi_r,[\chi_i,\chi_j]],
\en
\noi where
\eq
[\chi_i,\chi_j] \equiv \chi_i \chi_j - \Rhat{kl}{ij} \chi_k\chi_l
\en
\noi is the deformed commutator of eq. (\ref{qLie}).
\sk
viii) The left-invariant $\om^i$ satisfy the $q$-analogue of the
{\sl Cartan-Maurer equations}:
\eq
d\om^i+\c{jk}{i} \om^j \we \om^k=0, \label{CM}
\en
\noi where
\eq
\c{jk}{i} \equiv \chi_j \chi_k (x^i) \label{cijk}
\en
\eq
\chi_i(x^k)\equiv \de^k_i. \label{qcoord}
\en
The $x^k \in A$ defined in (\ref{qcoord}) are called the ``coordinates
of $G_q$" and satisfy $\epsi(x^i)=0$, which is the $q$-analogue of the
fact that classically they vanish at the origin of $G$ [recall that
$\epsi(x^i) \qonelim x^i(e)$]. Such $x^i$ can always be found
\cite{Wor}. Note that $\chi_i * x^j$ is the $q$-analogue of the inverse
vielbein.

The structure constants $C$ satisfy the Jacobi
identities obtained by taking
the exterior derivative of (\ref{CM}):
\eq
(\c{jk}{i} \c{rs}{j} - \c{rj}{i} \c{sk}{j} )\om^r\we\om^s\we\om^k=0.
\label{cJacobi}
\en
In the $q=1$ limit, $\om^j \we \om^k$ becomes antisymmetric
in {\sl j} and {\sl k}, and we have
\eq
\c{jk}{i} \qonelim =\unmezzo(\chi_j \chi_k - \chi_k\chi_j)(x^i)
=\unmezzo \C{jk}{l}~ \chi_l (x^i) = \unmezzo \C{jk}{i},
\en
\noi where $\C{jk}{l}$ are now the classical structure constants. Thus
when $q=1$ we have $\c{jk}{i}= \unmezzo \C{jk}{i}$ and (\ref{CM})
reproduces the classical Cartan-Maurer equations.

For $q \not= 1$, we find the following relation:
\eq
\C{jk}{i} = \c{jk}{i} - \Rhat{rs}{jk} \c{rs}{i} \label{Ccrelation}
\en
\noi after applying both members of eq. (\ref{qLie}) to $x^i$. Note
that, using (\ref{Ccrelation}), the Cartan-Maurer equations (\ref{CM})
can also be written as:
\eq
d\om^i+\C{jk}{i} \om^j \otimes \om^k=0.
\en
\sk
ix) Finally, we derive two operatorial identities that become trivial in
the limit $q \rightarrow 1$. From the formula
\eq
d(h * \theta)=h * d\theta,~~~h \in \Ap,~\theta \in \Ga^
{\we n}\label{dhthe}
\en
\noi [a direct consequence of (\ref{propd4})] with $h=\f{n}{l}$, we find
\eq
\chi_k  \f{n}{l}=\Rhat{ij}{kl} \f{n}{i} \chi_j  \label{bic4}
\en

By requiring consistency between the external derivative and the
bimodule structure of $\Ga$, i.e. requiring that
\eq
d(\om^i a)=d[(\f{i}{j} * a) \om^j],
\en
\noi one finds the identity
\eq
\C{mn}{i} \f{m}{j} \f{n}{k} + \f{i}{j} \chi_k= \Rhat{pq}{jk} \chi_p
\f{i}{q} + \C{jk}{l} \f{i}{l}. \label{bic3}
\en
\noi See Appendix A for the derivation of (\ref{bic4}) and
(\ref{bic3}).
\sk
In summary, a bicovariant calculus on a $G_q$ is characterized
by functionals $\chi_i$ and $\f{i}{j}$ on $A$
(``the algebra of functions on the quantum group") satisfying
\eq
\chi_i \chi_j - \Rhat{kl}{ij} \chi_k \chi_l = \C{ij}{k} \chi_k
\label{bico1}
\en
\eq
\Rhat{nm}{ij} \f{i}{p} \f{j}{q} = \f{n}{i} \f{m}{j} \Rhat{ij}{pq}
\label{bico2}
\en
\eq
\C{mn}{i} \f{m}{j} \f{n}{k} + \f{i}{j} \chi_k= \Rhat{pq}{jk} \chi_p
\f{i}{q} + \C{jk}{l} \f{i}{l} \label{bico3}
\en
\eq
\chi_k  \f{n}{l}=\Rhat{ij}{kl} \f{n}{i} \chi_j,  \label{bico4}
\en
\noi where the $q$-structure constants are given by $\C{jk}{i}=
\chi_k(\M{j}{i})$ and the braiding matrix by $\Rhat{ij}{kl}=
 \f{i}{l} (\M{k}{j})$.
In fact, these four relations seem to be also {\sl sufficient} to define
a bicovariant differential calculus on $A$ (see e.g. \cite{Bernard}).
By applying them to the element $\M{r}{s}$ we express these
relations (henceforth called {\sl bicovariance conditions})
in the adjoint representation:
\eqa
& & \C{ri}{n} \C{nj}{s}-\R{kl}{ij} \C{rk}{n} \C{nl}{s} =
\C{ij}{k} \C{rk}{s}
{}~~\mbox{({\sl q}-Jacobi identities)} \label{bicov1}\\
& & \Rhat{nm}{ij} \Rhat{ik}{rp} \Rhat{js}{kq}=\Rhat{nk}{ri} \Rhat{ms}{kj}
\Rhat{ij}{pq}~~~~~~~~~\mbox{(Yang--Baxter)} \label{bicov2}\\
& & \C{mn}{i} \Rhat{ml}{rj} \Rhat{ns}{lk} + \Rhat{il}{rj} \C{lk}{s} =
\Rhat{pq}{jk} \Rhat{il}{rq} \C{lp}{s} + \C{jk}{m} \Rhat{is}{rm}
\label{bicov3}\\
& & \C{rk}{m} \Rhat{ns}{ml} = \Rhat{ij}{kl} \Rhat{nm}{ri} \C{mj}{s}
\label{bicov4}
\ena
We conclude this section by giving the co-structures on the
quantum Lie algebra generators $\chi_i$ [those on the functionals
$\f{i}{j}$ have been given in (\ref{copf})-(\ref{coif})]:
\eqa
& & \Dp (\chi_i)=\chi_j
     \otimes \f{j}{i} + \Ip \otimes \chi_i \label{copchi}\\
& & \ep(\chi_i)=0 \label{couchi}\\
& & \kp(\chi_i)=-\chi_j \kp(\f{j}{i}), \label{coichi}
\ena
\noi which $q$-generalize the ones given in (\ref{copL})-(\ref{coiL}).
These co-structures derive from (\ref{copdual})-(\ref{coidual}). For
example, using (\ref{propd1}) and (\ref{omb}), eq. (\ref{copdual}) yields
(\ref{copchi}). They are consistent with the bicovariance
conditions (\ref{bico1})-(\ref{bico4}).
\sk
In the next section, we describe a constructive procedure due to
Jur\v co
\cite{Jurco} for a bicovariant differential calculus on any $q$-group of
the $A,B,C,D$ series considered in \cite{FRT}. The procedure is
illustrated on the example of $GL_q(2)$, for which all the objects
$\f{i}{j},~\M{r}{s},~\Rhat{ij}{kl},~\c{jk}{i}$ and $\C{jk}{i}$
are explicitly computed.

\sect{Constructive procedure and the example of $GL_q(2)$}

The generic $q$-group discussed in Section 3 is characterized by the
matrix $\R{ab}{cd}$. In terms of this matrix, it is possible to
construct a bicovariant differential calculus on the $q$-group. The
general procedure is described in this section, and the results for the
specific case of $GL_q(2)$ are collected in the table.
\sk
{\bf The $L^{\pm}$ functionals}
\sk
We start by introducing the linear functionals $\Lpm{a}{b}$, defined
by their value on the elements $\T{a}{b}$:
\eq
\Lpm{a}{b} (\T{c}{d})=\Rpm{ac}{bd}, \label{defL}
\en
\noi where
\eq
\Rp{ac}{bd} \equiv c^+ \R{ca}{db} \label{Rplus}
\en
\eq
\Rm{ac}{bd} \equiv c^- \Rinv{ac}{bd}, \label{Rminus}
\en
%\noi where $c=q^{{-1\over n}}$ for the series $A_{n-1}$ and $c=1$ for
%the $B,C,D$ series.
\noi where $c^+$, $c^-$ are free parameters (see later). The
inverse matrix $R^{-1}$ is defined by
\eq
\Rinv{ab}{cd}\R{cd}{ef} \equiv \de^
a_e \de^b_f \equiv \R{ab}{cd}\Rinv{cd}{ef}.
\en
\noi We see that the $\Lpm{a}{b}$ functionals are dual to
the $\T{a}{b}$ elements (fundamental representation) in the same way
the $\f{i}{j}$ functionals are dual to the $\M{i}{j}$ elements
of the adjoint representation. To extend the definition (\ref{defL})
to the whole algebra $A$, we set:
\eq
\Lpm{a}{b} (ab)=\Lpm{a}{g} (a) \Lpm{g}{b} (b),~~~\forall a,b\in A
\label{Lab}
\en
\noi so that, for example,
\eq
\Lpm{a}{b} (\T{c}{d} \T{e}{f}) = \Rpm{ac}{gd}\Rpm{ge}{bf}.
\en
\noi In general, using the compact notation introduced in Section 3,
\eq
\LLpm_1(T_2T_3...T_n)=\RRpm_{12} \RRpm_{13} ... \RRpm_{1n}. \label{LTT}
\en
\noi Finally, the value of $\LLpm$ on the unit $I$ is defined by
\eq
\Lpm{a}{b} (I)=\de^a_b. \label{LI}
\en
Thus the functionals $\Lpm{a}{b}$ have the same properties as their
adjoint counterpart $\f{i}{j}$, and not surprisingly the latter will be
constructed in terms of the former.

{}From (\ref{LTT}) we can also find the action of $\Lpm{a}{b}$ on $a\in A
$, i.e. $\Lpm{a}{b} * a$. Indeed
\eqa
\lefteqn{\Lpm{a}{b} * (\T{c_1}{d_1} \T{c_2}{d_2}\cdots\T{c_n}{d_n})=
 [id \otimes \Lpm{a}{b}] \D(\T{c_1}{d_1} \T{c_2}{d_2} \cdots
\T{c_n}{d_n})=}
 \nonumber\\
& & [id \otimes \Lpm{a}{b}]\D(\T{c_1}{d_1})\cdots\D(\T{c_n}{d_n})=
 \nonumber\\
& & [id\otimes \Lpm{a}{b}] (\T{c_1}{e_1}\cdots\T{c_n}{e_n}\otimes
  \T{e_1}{d_1}\cdots\T{e_n}{d_n})\nonumber\\
& & \T{c_1}{e_1}\cdots\T{c_n}{e_n}\Lpm{a}{b}(\T{e_1}{d_1}\cdots
\T{e_n}{d_n})=\nonumber \\
& & \T{c_1}{e_1}\cdots\T{c_n}{e_n} \Rpm{ae_1}{g_1d_1}
\Rpm{g_1e_2}{g_2d_2}\cdots\Rpm{g_{n-1}e_n}{~~~bd_n}
\ena
\noi or, more compactly,
\eq
\LLpm_1 * T_2...T_n=T_2...T_n \RRpm_{12} \RRpm_{13} ...\RRpm_{1n},
\en
\noi which can also be written as
\eq
\LLpm_1 * T_2=T_2 \RRpm_{12}\LLpm_1.
\en

It is not difficult to find the commutations between $\Lpm{a}{b}$
and $\Lpm{c}{d}$:
\eq
R_{12} \LLpm_2 \LLpm_1=\LLpm_1 \LLpm_2 R_{12} \label{RLL}
\en
\eq
R_{12} \LLp_2 \LLm_1=\LLm_1 \LLp_2 R_{12}, \label{RLpLm}
\en
\noi where as usual the product $\LLpm_2 \LLpm_1$ is the convolution
product $\LLpm_2 \LLpm_1 \equiv (\LLpm_2 \otimes \LLpm_1)\D$. Consider
\eq
R_{12} (\LLp_2 \LLp_1)(T_3)=R_{12}(\LLp_2 \otimes\LLp_1)\D (T_3)=
R_{12}(\LLp_2 \otimes\LLp_1)(T_3 \otimes T_3)=(c^+)^2~R_{12}R_{32}R_{31}
\nonumber
\en
\noi and
\eq
\LLp_1 \LLp_2(T_3) R_{12}=(c^+)^2 ~R_{31}R_{32}R_{12}
\nonumber
\en
\noi so that the equation (\ref{RLL}) is proven for $\LLp$
by virtue of the
quantum Yang--Baxter equation (\ref{YB}), where the indices have been
renamed $2\rightarrow 1,3 \rightarrow 2,1\rightarrow 3$. Similarly,
one proves the remaining ``RLL" relations.
\sk
{\sl Note 1 :} As mentioned in \cite{FRT}, $L^+$ is upper
triangular, $L^-$ is lower triangular (this is due to the upper
and lower
triangularity of $R^+$ and $R^-$, respectively).
\sk
{\sl Note 2 :} When $\det_q T=1$, we have $det_{q^{-1}} L^{\pm}=\Lpm{1}{1}
\Lpm{2}{2}
\cdots \Lpm{n}{n}=\epsi$, and $\Lp{i}{i} \Lm{j}{j}=\Lm{j}{j} \Lp{i}{i}$
(no sum on repeated indices). Then $(det_{q^{-1}} L^{\pm}) (\det_q T)=
\det_q R^{\pm}$ implies $det R^{\pm}=1$, which requires
$c^+=q^{-1/n}$, $c^-=q^{1/n}$ for the $A_{n-1}$ series (and $c^{\pm}=1$
for the remaining $B,C,D$ series) in (\ref{Rplus}) and (\ref{Rminus}).
In the more general case of
$GL_q(n)$, $c^{\pm}$ are extra free parameters, cf. \cite{Sun}.
In fact, they appear only in the combination $s=(c^+)^{-1} c^-$.
They do not enter in the $\Rh$ matrix, nor in the structure
constants or the Cartan-Maurer
equations (see the table). Different values
of $s$ lead to isomorphic differential calculi (in the sense of ref.
\cite{Wor}), so that $s$ is not really an essential parameter.
\sk
The co-structures are defined by the duality (\ref{defL}):
\eq
\Dp(\Lpm{a}{b})(\T{c}{d} \otimes \T{e}{f}) \equiv \Lpm{a}{b}
(\T{c}{d}\T{e}{f})=\Lpm{a}{g}(\T{c}{d}) \Lpm{g}{b} (\T{e}{f})
\en
\eqa
& & \ep (\Lpm{a}{b})\equiv \Lpm{a}{b} (I)\\
& & \kp (\Lpm{a}{b})(\T{c}{d})\equiv \Lpm{a}{b} (\kappa (\T{c}{d}))
\ena
\noi cf. [(\ref{copdual})-(\ref{coidual})], so that
\eqa
& & \Dp (\Lpm{a}{b})=\Lpm{a}{g} \otimes \Lpm{g}{b}\label{copLpm}\\
& & \ep (\Lpm{a}{b})=\de^a_b \label{couLpm}\\
& & \kp (\Lpm{a}{b})=\Lpm{a}{b} \circ \kappa \label{coiLpm}
\ena
\noi and the $\Lpm{a}{b}$ generate the Hopf algebra dual to the quantum
group. Note that
\eq
\Lpm{a}{b} (\kappa (\T{c}{d}))=\Rpminv{ac}{bd}, \label{LkT}
\en
\noi since
\eq
\Lpm{a}{b} (\kappa (\T{c}{d}) \T{d}{e} )=\de^c_d \Lpm{a}{b} (I)=
\de^c_e \de^a_b
\en
\noi and
\eqa
\Lpm{a}{b} (\kappa (\T{c}{d}) \T{d}{e})&=& \Lpm{a}{f} (\kappa(\T{c}{d}))
\Lpm{f}{b} (\T{d}{e}) \nonumber\\
&=&\Lpm{a}{f} (\kappa (\T{c}{d})) \Rpm{fd}{be}.
\ena
\sk
{\bf The space of quantum 1-forms}
\sk
The bimodule $\Ga$ (``space of quantum 1-forms") can be constructed as
follows. First we define $\ome{a}{b}$ to be a basis of left-invariant
quantum 1-forms. The index pairs ${}_a^{~b}$ or ${}^a_{~b}$
will replace in the
sequel the indices ${}^i$ or ${}_i$ of the previous section. The
dimension of $\invG$ is therefore $N^2$ at this stage. The existence of
this basis can be proven by considering $\Ga$ to be the tensor product
of two fundamental bimodules, see refs. \cite{Jurco,Watamura}.
Here we just assume that it exists. Since the $\ome{a}{b}$ are
left-invariant, we have:
\eq
\DL (\ome{a}{b})=I\otimes \ome{a}{b},~~~a,b=1,...,N. \label{DLome}
\en
\noi The left action $\DL$ on the whole of $\Ga$ is then defined by
(\ref{DLome}), since $\ome{a}{b}$ is a basis for $\Ga$. The bimodule
$\Ga$ is further characterized by the commutations
between $\ome{a}{b}$ and $a \in A$ [cf. eq. (\ref{omb})]:
\eq
\ome{a_1}{a_2} b=(\ff{a_1}{a_2b_1}{b_2} * b)\ome{b_1}{b_2}, \label{omeb}
\en
\noi where
\eq
\ff{a_1}{a_2b_1}{b_2} \equiv \kp (\Lp{b_1}{a_1}) \Lm{a_2}{b_2}.
\label{defff}
\en
\noi Finally, the right action $\DR$ on $\Ga$ is defined by
\eq
\DR (\ome{a_1}{a_2}) = \ome{b_1}{b_2} \otimes \MM{b_1}{b_2a_1}{a_2},
\en
\noi where $\MM{b_1}{b_2a_1}{a_2}$, the adjoint representation,
is given by
\eq
\MM{b_1}{b_2a_1}{a_2} \equiv \T{b_1}{a_1} \kappa (\T{a_2}{b_2}).
\label{defMM}
\en
It is easy to check that $\ff{a_1}{a_2b_1}{b_2}$ fulfill the
consistency conditions (\ref{propf1})-(\ref{propf3}), where
the {\small{\sl i,j,...}} indices stand for pairs of {\small{\sl a,b,...}
} indices. Also, the
co-structures
are as given in (\ref{copf})-(\ref{coif}).
\sk
{\bf The $\Rh$ tensor and the exterior product}
\sk
The $\Rh$ tensor defined in (\ref{defR}) can now be computed:
\eqa
\lefteqn{\RRhat{a_1}{a_2}{d_1}{d_2}{c_1}{c_2}{b_1}{b_2}
\equiv \ff{a_1}{a_2b_1}{b_2} (\MM{c_1}{c_2d_1}{d_2})
=\kp(\Lp{b_1}{a_1}) \Lm{a_2}{b_2} (\T{c_1}{d_1} \kappa(\T{d_2}{c_2}))
}\nonumber \\
=& & [\kp (\Lp{b_1}{a_1}) \otimes \Lm{a_2}{b_2}]\D (\T{c_1}{d_1} \kappa
(\T{d_2}{c_2}))\nonumber\\
=& & [\kp (\Lp{b_1}{a_1}) \otimes \Lm{a_2}{b_2}]
(\T{c_1}{e_1} \otimes \T{e_1}{d_1})(\kappa(\T{f_2}{c_2})\otimes \kappa
(\T{d_2}{f_2})) \nonumber \\
=& & [\kp (\Lp{b_1}{a_1}) \otimes \Lm{a_2}{b_2}]
[\T{c_1}{e_1}\kappa(\T{f_2}{c_2}) \otimes \T{e_1}{d_1}\kappa
(\T{d_2}{f_2})] \nonumber\\
=& & \Lp{b_1}{a_1} (\kappa^2 (\T{f_2}{c_2}) \kappa (\T{c_1}{e_1}))
{}~\Lm{a_2}{b_2} (\T{e_1}{d_1} \kappa (\T{d_2}{f_2}))\nonumber \\
=& & d^{f_2} d^{-1}_{c_2} \Lp{b_1}{a_1} (\T{f_2}{c_2} \kappa (\T{c_1}{e_1
}))~\Lm{a_2}{b_2} (\T{e_1}{d_1} \kappa(\T{d_2}{f_2})\nonumber \\
=& & d^{f_2} d^{-1}_{c_2} \Lp{b_1}{g_1} (\T{f_2}{c_2}) ~\Lp{g_1}{a_1}
(\kappa (\T{c_1}{e_1}))~ \Lm{a_2}{g_2} (\T{e_1}{d_1}) ~\Lm{g_2}{b_2}
(\kappa(\T{d_2}{f_2}))\nonumber \\
=& & d^{f_2} d^{-1}_{c_2} \R{f_2b_1}{c_2g_1} \Rinv{c_1g_1}{e_1a_1}
    \Rinv{a_2e_1}{g_2d_1} \R{g_2d_2}{b_2f_2} \label{RRffMM}
\ena
\noi where we made use of relations (\ref{Dka}), (\ref{coidual}),
(\ref{k2}), (\ref{defL}) and (\ref{LkT}). The
$\Rh$ tensor allows the definition of the exterior product as in
(\ref{exom}). For future use we give here also the inverse $\Rh^{-1}$
of the $\Rh$ tensor, defined by:
\eq
\RRhatinv{a_1}{a_2}{d_1}{d_2}{b_1}{b_2}{c_1}{c_2} \RRhat{b_1}{b_2}{c_1}
{c_2}{e_1}{e_2}{f_1}{f_2}= \de^{a_2}_{e_2} \de^{e_1}_{a_1} \de^{f_1}_
{d_1} \de^{d_2}_{f_2}. \label{defRRhatinv}
\en
It is not difficult to see that
\eqa
\lefteqn{\RRhatinv{a_1}{a_2}{d_1}{d_2}{b_1}{b_2}{c_1}{c_2}=
\ff{d_1}{d_2b_1}{b_2}
(\T{a_2}{c_2} \km (\T{c_1}{a_1}))=}\nonumber\\
& & \R{f_1b_1}{a_1g_1} \Rinv{a_2g_1}{e_2d_1} \Rinv{d_2e_2}{g_2
c_2} \R{g_2c_1}{b_2f_1} (d^{-1})^{c_1} d_{f_1}
\label{RRffMMinv}
\ena
\noi does the trick. Another useful relation gives a particular
trace of the $\Rh$ matrix:
\eq
\RRhat{c_1}{c_2}{b}{b}{a_1}{a_2}{b_1}{b_2} = \de^{a_1}_{a_2}
\de^{b_1}_{c_1} \de^{c_2}_{b_2}. \label{RReqdeltas}
\en
\noi This identity is simply proven. Indeed:
\eqa
\lefteqn{\RRhat{c_1}{c_2}{b}{b}{a_1}{a_2}{b_1}{b_2}
\equiv \ff{c_1}{c_2b_1}{b_2} (\MM{a_1}{a_2 b}{b})=}\nonumber\\
& & \kp(\Lp{b_1}{c_1}) \Lm{c_2}{b_2} (\T{a_1}{b} \kappa(\T{b}{a_2}))=
\kp(\Lp{b_1}{c_1}) \Lm{c_2}{b_2} (\de^{a_1}_{a_2} I)=\nonumber\\
& & \de^{a_1}_{a_2} [\kp(\Lp{b_1}{c_1}) \otimes \Lm{c_2}{b_2}]
    (I\otimes I)=\de^{a_1}_{a_2}\de^{b_1}_{c_1} \de^{c_2}_{b_2}.
\ena
\sk
The relations (\ref{Hecke}), (\ref{R3}) for the $R$ matrix
reflect themselves in relations
for the $\Rh$ matrix (\ref{RRffMM}). For example, the Hecke condition
(\ref{Hecke}) implies:
\eq
(\Rh + q^2)(\Rh + q^{-2})(\Rh-I)=0 \label{RRHecke}
\en
\noi for the $A_{n-1}$ $q$-groups, and replaces the
classical relation $(\Rh -1)(\Rh+1)=0$, $\Rh$ being for $q=1$ the
ordinary permutation operator, cf. (\ref{limRhat}).

With the help of
(\ref{RRHecke}) we can give explicitly the commutations of the
left-invariant forms $\om$. Indeed, reverting to the
{\small{\sl i,j...}} indices,
relation (\ref{RRHecke}) implies:
\eqa
\lefteqn{(\Rhat{ij}{kl} + q^2 \de^{i}_{k} \de^{j}_{l})
(\Rhat{kl}{mn} + q^{-2} \de^{k}_{m} \de^{l}_{n})
(\Rhat{mn}{rs} -  \de^{m}_{r} \de^{n}_{s})\om^r \otimes \om^s=}
\nonumber\\
& & (\Rhat{ij}{kl} + q^2 \de^{i}_{k} \de^{j}_{l})
(\Rhat{kl}{mn} + q^{-2} \de^{k}_{m} \de^{l}_{n})
\om^m \we \om^n = 0
\ena
\noi and it is easy to see that the last equality can be rewritten as
\eq
\om^i \we \om^j = - \Z{ij}{kl} \om^k \we \om^l \label{commom}
\en
\eq
\Z{ij}{kl} \equiv {1\over {q^2 + q^{-2}}} [\Rhat{ij}{kl} + \Rhatinv
{ij}{kl}]. \label{defZ}
\en
\sk
{\bf The exterior differential}
\sk
The exterior differential on $\Ga^{\we k}$ is defined by means of the
bi-invariant (i.e. left- and right-invariant) element $\tau=\sum_a
\ome{a}{a} \in \Ga$ as follows:
\eq
d\theta \equiv \lam [\tau \we \theta - (-1)^k \theta \we \tau],
\label{defd}
\en
\noi where $\theta \in \Ga^{\we k}$, and $\la$ is a
normalization factor depending on $q$, necessary in order to obtain the
correct classical limit. It will be later determined to be
$\la = q-q^{-1}$. Here we can only see that it has to vanish for $q=1$,
since otherwise $d\theta$ would vanish in the classical limit. For
$a \in A$ we have
\eq
da=\lam [\tau a - a \tau]. \label{defd2}
\en
\noi This linear map satisfies the Leibniz rule (\ref{Leibniz}), and
properties
(\ref{propd1})-(\ref{propd4}), as the reader can easily check (use the
definition of exterior product and the bi-invariance of $\tau$). A proof
that also the property (\ref{adb}) holds can be
obtained by considering the
exterior differential of the adjoint representation:
\eq
d\M{j}{i}=(\chi_k * \M{j}{i} )\om^k=\M{j}{l} \C{kl}{i}\om^k
\en
\noi or
\eq
\kappa (\M{l}{j})d\M{j}{i}=\C{kl}{i} \om^k.
\en
\noi Multiplying by $\C{ni}{l}$, we have:
\eq
\C{ni}{l} \kappa (\M{l}{j}) d\M{j}{i}=\C{kl}{i} \C{ni}{l}\om^k\equiv
g_{nk} \om^k,
\en
\noi where $g_{nk}$ is the $q$-Killing metric.
%For $q=1$
%semisimple Lie groups this metric is invertible, and so is its
%continuous deformation, at least in the proximity of $q=1$.
The explicit example of this paper being $GL_q (2)$, one may wonder
what happens to the invertibility of the $q$-Killing metric, since its
classical limit is no more invertible [$GL(2)$ being nonsemisimple].
The answer is that for $q \not= 1$ the $q$-Killing metric of $GL_q(2)$
{\sl is} invertible, as can be checked explicitly from the values of the
structure constants given in the table. Therefore $GL_q(2)$
could be said to
be ``$q$-semisimple". With an analogous procedure (using $\T{a}{b}$
instead of $\M{j}{i}$) we have derived in the table the
explicit expression of the
$\om^i$ in terms of the $d\T{a}{b}$ for $GL_q(2)$.
\sk
{\bf The $q$-Lie algebra}
\sk
The ``quantum generators" $\cchi{a_1}{a_2}$ are introduced as in
(\ref{defchi}):
\eq
da=\lam[\tau a - a\tau] =(\cchi{a_1}{a_2} * a)
\ome{a_1}{a_2}. \label{defcchi}
\en
Using (\ref{omeb}) we can find an explicit expression for the
$\cchi{a_1}{a_2}$ in terms of the $\LLpm$
functionals. Indeed
\eq
\tau a= \ome{b}{b} a= (\ff{b}{bc_1}{c_2} * a)\ome{c_1}{c_2}=
([\kp (\Lp{c_1}{b}) \Lm{b}{c_2}] * a) \ome{c_1}{c_2}.
\en
\noi Therefore
\eq
da=\lam [(\kp (\Lp{c_1}{b}) \Lm{b}{c_2}- \de^{c_1}_{c_2} \epsi) * a]
\ome{c_1}{c_2} \label{explicitda}
\en
\noi (recall $\epsi * a= a$), so that the $q$-generators take the
explicit form
\eq
\cchi{c_1}{c_2}=\lam [\kp (\Lp{c_1}{b})\Lm{b}{c_2}-\de^{c_1}_{c_2}
\epsi ].
\label{defchi2}
\en
\noi The commutations between the $\chi$'s can now be obtained by taking
the exterior derivative of eq. (\ref{explicitda}). We find
\eqa
\lefteqn{d^2(a)=0=d[(\cchi{c_1}{c_2} * a) \ome{c_1}{c_2}]=
(\cchi{d_1}{d_2}*\cchi{c_1}{c_2} * a) \ome{d_1}{d_2} \we \ome{c_1}{c_2}
 + (\cchi{c_1}{c_2} * a)d\ome{c_1}{c_2} }\nonumber\\
& & =(\cchi{d_1}{d_2}*\cchi{c_1}{c_2} * a) (\ome{d_1}{d_2} \otimes
\ome{c_1}{c_2}- \RRhat{d_1}{d_2}{c_1}{c_2}{e_1}{e_2}{f_1}{f_2}
\ome{e_1}{e_2} \otimes \ome{f_1}{f_2})  \nonumber\\
& & +\lam (\cchi{c_1}{c_2} * a)
(\ome{b}{b} \we \ome{c_1}{c_2} + \ome{c_1}{c_2} \we \ome{b}{b}).
\label{d2}
\ena
\noi Now we use the fact that $\tau=\ome{b}{b}$ is bi-invariant, and
therefore also right-invariant, so that we can write
\eqa
\lefteqn{\ome{b}{b} \we \ome{c_1}{c_2} + \ome{c_1}{c_2} \we \ome{b}{b}
\equiv}\nonumber\\
& & \ome{b}{b} \otimes \ome{c_1}{c_2} - \Rh (\ome{b}{b} \otimes
\ome{c_1}{c_2})+\ome{c_1}{c_2} \otimes \ome{b}{b} -
\Rh (\ome{c_1}{c_2} \otimes \ome{b}{b})=\nonumber\\
& & \ome{c_1}{c_2} \otimes \ome{b}{b} - \Rh (\ome{b}{b} \otimes
\ome{c_1}{c_2})=\nonumber\\
& & \ome{c_1}{c_2} \otimes \ome{b}{b} - \RRhat{b}{b}{c_1}{c_2}{e_1}{e_2
}{f_1}{f_2} \ome{e_1}{e_2} \otimes \ome{f_1}{f_2},
\label{om2}
\ena
\noi where we have used $\Rh(\ome{c_1}{c_2} \otimes \tau) =
\tau \otimes \ome{c_1}{c_2}$, cf. (\ref{Rhat}). After substituting
(\ref{om2}) in (\ref{d2}), and factorizing $\ome{d_1}{d_2} \otimes
\ome{c_1}{c_2}$, we arrive at the $q$-Lie algebra relations:
\eq
\cchi{d_1}{d_2} \cchi{c_1}{c_2} - \RRhat{e_1}{e_2}{f_1}{f_2}
{d_1}{d_2}{c_1}{c_2} ~\cchi{e_1}{e_2} \cchi{f_1}{f_2} =
\lam [-\de^{c_1}_{c_2} \cchi{d_1}{d_2} + \RRhat{b}{b}{e_1}{e_2}{d_1}{d_2}
{c_1}{c_2} ~\cchi{e_1}{e_2}].
\label{qLieexplicit}
\en
\noi The structure constants are then explicitly given by:
\eq
\CC{a_1}{a_2}{b_1}{b_2}{c_1}{c_2} =\lam [- \de^{b_1}_{b_2}
\de^{a_1}_{c_1}
\de^{c_2}_{a_2} + \RRhat{b}{b}{c_1}{c_2}{a_1}{a_2}{b_1}{b_2}]. \label{CC}
\en
Here we determine $\la$. Indeed we first observe that
\eq
\RRhat{a_1}{a_2}{d_1}{d_2}{c_1}{c_2}{b_1}{b_2}=
\de^{b_1}_{a_1} \de^{a_2}_{b_2} \de^{c_1}_{d_1} \de^{d_2}_{c_2}
+ (q-q^{-1}) \U{a_1}{a_2}{d_1}{d_2}{c_1}{c_2}{b_1}{b_2},
  \label{Rlam}
\en
\noi where the matrix $U$ is finite and different from zero in the limit
$q=1$. This can be proven by considering the explicit form of the
$R$ and $R^{-1}$ matrices. In the case of the $A_{n-1}$
$q$-groups, for example, these matrices have the form \cite{FRT}:
\eq
\R{ab}{cd}= \de^{a}_{c} \de^b_d + (q-q^{-1}) \left[{{q-1}\over{q-q^{-1}}}
\de^a_c \de^b_d \de^{ab} + \de^b_c \de^a_d \theta (a-b) \right]
\label{explicitR}
\en
\eq
\Rinv{ab}{cd}= \de^{a}_{c} \de^b_d - (q-q^{-1}) \left[{{1-q^{-1}}
\over{q-q^{-1}}}
\de^a_c \de^b_d \de^{ab} + \de^b_c \de^a_d \theta (a-b) \right],
\label{explicitRinv}
\en
\noi where $\theta (x) = 1$ for $x > 0$ and vanishes for $x \leq 0$.
Substituting these expressions in the formula for $\Rh$ (\ref{RRffMM})
we find (\ref{Rlam}). Using (\ref{Rlam}) in the expression
(\ref{CC}) for the $q$-structure constants $\Cb$, we find that the
terms proportional to $\lam$ do cancel, and we are left with:
\eq
\CC{a_1}{a_2}{b_1}{b_2}{c_1}{c_2} =- \lam (q-q^{-1})
\U{b}{b}{c_1}{c_2}{a_1}{a_2}{b_1}{b_2}. \label{CClam}
\en
\noi A simple choice for $\la$ is therefore $\la=q-q^{-1}$, ensuring
that $\Cb$ remains finite in the limit \qone.
\sk
{\bf The Cartan-Maurer equations}
\sk
The Cartan-Maurer equations are found as follows:
\eq
d\ome{c_1}{c_2}=\lam (\ome{b}{b} \we \ome{c_1}{c_2} + \ome{c_1}{c_2} \we
\ome{b}{b}) \equiv -\cc{a_1}{a_2}{b_1}{b_2}{c_1}{c_2} ~\ome{a_1}{a_2} \we
\ome{b_1}{b_2}. \label{CartanMaurer}
\en
In order to obtain an explicit expression for the $C$ structure
constants in (\ref{CartanMaurer}), we must use the relation
(\ref{commom}) for the commutations of $\ome{a_1}{a_2}$ with
$\ome{b_1}{b_2}$. Then the term $\ome{c_1}{c_2} \we \ome{b}{b}$
in (\ref{CartanMaurer}) can be written as
$-Z\om\om$ via formula (\ref{commom}), and we find the
$C$-structure constants to be:
\eqa
\cc{a_1}{a_2}{b_1}{b_2}{c_1}{c_2}&=&
-\lam(\de^{a_1}_{a_2} \de^{b_1}_{c_1}
\de^{c_2}_{b_2} - {1 \over {q^2 + q^{-2}}}[ \RRhat{c_1}{c_2}{b}{b}{a_1}
{a_2}{b_1}{b_2}+\RRhatinv{c_1}{c_2}{b}{b}{a_1}{a_2}{b_1}{b_2}])
\nonumber\\
&=& -\lam(\de^{a_1}_{a_2} \de^{b_1}_{c_1}
\de^{c_2}_{b_2} - {1 \over {q^2 + q^{-2}}}[ \de^{a_1}_{a_2}
\de^{b_1}_{c_1}\de^{c_2}_{b_2}+
\RRhatinv{c_1}{c_2}{b}{b}{a_1}{a_2}{b_1}{b_2}]),
\label{explicitcc}
\ena
\noi where we have also used eq. (\ref{RReqdeltas}). By considering
the analogue of (\ref{Rlam}) for $\Rh^{-1}$,
it is not difficult to see that the terms proportional to $\lam$
cancel, and the \qone limit of (\ref{explicitcc}) is well defined.
For a more detailed discussion, including also the $B_n,C_n$ and
$D_n$ $q$-groups, we refer to \cite{AC1}.
\sk
In the table we summarize the results of this section for the
case of $GL_q(2)$. The composite indices ${}_a^{~b}$ are translated into
the corresponding indices ${}^i$, $i=1,+,-,2$, according to
the convention:
\eq
{}_1^{~1} \rightarrow {}^1,~~{}_1^{~2} \rightarrow {}^+,~~{}_2^{~1}
\rightarrow {}^-,~~{}_2^{~2} \rightarrow {}^2.
\en
\noi A similar convention holds for ${}^a_{~b} \rightarrow {}_i$.
\sk

%As a check, we can verify that the relation (\ref{Ccrelation}) holds.
%In our composite index notation, this reads:
%\eq
%\CC{a_1}{a_2}{b_1}{b_2}{c_1}{c_2}=\cc{a_1}{a_2}{b_1}{b_2}{c_1}{c_2} -
%\RRhat{e_1}{e_2}{f_1}{f_2}{a_1}{a_2}{b_1}{b_2} \cc{e_1}{e_2}{f_1}{f_2
%}{c_1}{c_2}  \label{CCccrelation}
%\en
%\noi After substituting in this equation the explicit $\Cb$ and
%$C$ constants
%given in (\ref{CC}) and (\ref{cc}), we see that
%(\ref{CCccrelation}) holds if
%\eq
%\sk
\vfill\eject
\centerline{{\bf Table}}
\sk
\centerline{The bicovariant $GL_q(2)$ algebra}
\sk
\sk
\noi {\sl $R$ and $D$-matrices:}
\[
\R{ab}{cd}=\left( \begin{array}{cccc} q & 0 & 0 & 0 \\
                                      0 & 1 & 0 & 0 \\
                                      0 & q-\qm & 1 & 0 \\
                                      0 & 0 & 0 & q  \end{array} \right)
\]
\[
\Rm{ab}{cd}\equiv c^- \Rinv{ab}{cd}= c^- \left(
                   \begin{array}{cccc} \qm & 0 & 0 & 0 \\
                                      0 & 1 & 0 & 0 \\
                                      0 & -(q-\qm) & 1 & 0 \\
                                     0 & 0 & 0 & \qm  \end{array} \right)
\]
\[
\Rp{ab}{cd}\equiv c^+\R{ba}{dc}=c^+
                                      \left( \begin{array}{cccc}
                                      q & 0 & 0 & 0 \\
                                      0 & 1 & q-\qm & 0 \\
                                      0 & 0 & 1 & 0 \\
                                      0 & 0 & 0 & q  \end{array} \right)
,~~D^a_{~b}=\left( \begin{array}{cc} q & 0  \\ 0 & q^3 \end{array}
\right)
\]
\sk

\noi {\sl Non-vanishing components of the $\Rh$ matrix:}

\[
\begin{array}{llll}
\Rhat{11}{11}=1 &\Rhat{1+}{+1}=q^{-2} &\Rhat{1-}{-1}=q^2
&\Rhat{12}{21}=1\\
\Rhat{+1}{1+}=1 &\Rhat{+1}{+1}=1-q^{-2}
&\Rhat{++}{++}=1 &\Rhat{+-}{11}=1-q^2 \\
\Rhat{+-}{-+}=1 &\Rhat{+-}{21}=1-q^{-2}
&\Rhat{+2}{+1}=-1+q^{-2} &\Rhat{+2}{2+}=1\\
\Rhat{-1}{1-}=1 &\Rhat{-1}{-1}=1-q^{2}
&\Rhat{-+}{11}=-1+q^2 &\Rhat{-+}{+-}=1\\
\Rhat{-+}{21}=-1+q^{-2} &\Rhat{--}{--}=1
&\Rhat{-2}{-1}=-1+q^2 &\Rhat{-2}{2-}=1 \\
\Rhat{21}{11}=(q^2-1)^2 &\Rhat{21}{12}=1
&\Rhat{21}{+-}=q^2-1 &\Rhat{21}{-+}=1-q^2 \\
\Rhat{21}{21}=2-q^2-q^{-2} &\Rhat{2+}{1+}=-q^2+q^4
&\Rhat{2+}{+2}=q^2 &\Rhat{2+}{2+}=1-q^2\\
\Rhat{2-}{1-}=1-q^2 &\Rhat{2-}{-1}=q^{-2}-1-q^2+q^4
&\Rhat{2-}{-2}=q^{-2} &\Rhat{2-}{2-}=1-q^{-2}\\
\Rhat{22}{11}=-(q^2-1)^2 &\Rhat{22}{+-}=1-q^2
&\Rhat{22}{-+}=q^2-1 &\Rhat{22}{21}=(q^{-1}-q)^2\\
\Rhat{22}{22}=1& & &  \end{array}
\]
\sk
\noi {\sl Non-vanishing components of the $\Cb$ structure constants:}

\[
\begin{array}{llll}
\C{11}{1}=q(q^2-1) &\C{11}{2}=-q(q^2-1) &\C{1+}{+}=q^3
 &\C{1-}{-}=-q\\
\C{21}{1}=\qm-q &\C{21}{2}=q-\qm &\C{2+}{+}=-q
 &\C{2-}{-}=\qm\\
\C{+1}{+}=-\qm &\C{+2}{+}=q &\C{+-}{1}=q
  &\C{+-}{2}=-q\\
\C{-1}{-}=q(q^2+1)-\qm &\C{-2}{-}=-\qm &\C{-+}{1}=-q
  &\C{-+}{2}=q \end{array}
\]
\sk
\renewcommand{\baselinestretch}{1.2}
\noi {\sl Non-vanishing components of the $C$ structure constants:}
\[
\begin{array}{llll}
\c{11}{1}=\frac{q(q^2-1)^2}{1+q^4} &\c{11}{2}=\frac{q^3 (1-q^2)}
{1+q^4} &\c{1+}{+}=\frac{q^5}{1+q^4}
 &\c{1-}{-}=\frac{-q^3}{1+q^4}\\
\c{12}{1}=\frac{q(1-q^2)}{1+q^4}  &\c{+1}{+}=\frac{-q^3}{1+q^4}
&\c{+-}{1}=\frac{q^3}{1+q^4} &\c{+-}{2}= \frac{-q^3}{1+q^4}\\
\c{+2}{+}=\frac{q}{1+q^4} &\c{-1}{-}=\frac{q^5}{1+q^4}
&\c{-+}{1}=\frac{-q^3}{1+q^4}  &\c{-+}{2}=\frac{q^3}{1+q^4}\\
\c{-2}{-}=\frac{-q^3}{1+q^4} &\c{21}{1}=\frac{q(1-q^2)}{1+q^4}
&\c{2+}{+}=\frac{-q^3}{1+q^4} &\c{2-}{-}=\frac{q}{1+q^4}\\
\c{22}{2}=\frac{q(1-q^2)}{1+q^4} & & & \end{array}
\]
\sk
\renewcommand{\baselinestretch}{1.0}
\noi {\sl Cartan-Maurer equations:}
\[ d\om^1+q\om^+ \we \om^-=0 \]
\[ d\om^+ + q \om^+(-q^2 \om^1 + \om^2)=0 \]
\[ d\om^- + q (-q^2 \om^1 + \om^2)\om^-=0 \]
\[ d\om^2 - q \om^+ \we \om^-=0 \]
\sk

\noi {\sl The q-Lie algebra:}
\sk
\[ \chi_1 \chi_+ - \chi_+ \chi_1 -(q^4-q^2)\chi_2 \chi_+ = q^3 \chi_+\]
\[ \chi_1 \chi_- - \chi_- \chi_1 +(q^2-1)\chi_2 \chi_- = -q \chi_-\]
\[ \chi_1 \chi_2 - \chi_2 \chi_1 =0\]
\[ \chi_+ \chi_- - \chi_- \chi_+ + (1-q^2) \chi_2 \chi_1-(1-q^2)
\chi_2 \chi_2 = q (\chi_1-\chi_2)
\]
\[ \chi_+ \chi_2-q^2 \chi_2 \chi_+ = q\chi_+\]
\[ \chi_- \chi_2-q^{-2} \chi_2 \chi_- = -\qm \chi_-\]
\sk
\noi {\sl Commutation
relations between left-invariant $\om^i$ and $\om^j$:}
\[\om^1 \we \om^+ + \om^+ \we \om^1 = 0\]
\[\om^1 \we \om^- + \om^- \we \om^1 = 0\]
\[\om^1 \we \om^2 + \om^2 \we \om^1 = (1-q^2)\om^+ \we \om^-
\]
\[\om^+ \we \om^- + \om^- \we \om^+ = 0\]
\[\om^2 \we \om^+ + q^2 \om^+ \we \om^2 = q^2 (q^2 - 1)\om^+ \we \om^1
\]
\[\om^2 \we \om^- + q^{-2} \om^- \we \om^2 = (1-q^2)\om^- \we \om^1\]
\[\om^2 \we \om^2 =(q^2 - 1)\om^+ \we \om^-
\]
\[\om^1 \we \om^1 = \om^+ \we \om^+ = \om^- \we \om^-=0\]
\sk

\noi {\sl Commutation relations
between $\om^i$ and the basic elements of $A$ ($s=(c^+)^{-1} c^-$):}

\[
\begin{array}{ll}
\om^1 \al= sq^{-2} \al \om^1
& \om^+ \al= s\qm\al \om^+\\
\om^1\be=s\be\om^1 & \om^+\be=s\qm\be\om^+ + s(q^{-2} -1)\al\om^1\\
\om^1\ga=sq^{-2}\ga\om^1 & \om^+\ga=s\qm\ga\om^+\\
\om^1\de=s\de\om^1 & \om^+\de=s\qm\de\om^+ + s(q^{-2} -1)\ga\om^1
\end{array}
\]
\[
\begin{array}{ll}
\om^-\al=s\qm\al\om^- + s(q^{-2}-1)\be\om^1 & \om^2\al=s\al\om^2+s
(\qm-q)\be\om^+\\
\om^-\be=s\qm\be\om^- & \om^2\be=sq^{-2}\be\om^2+s(\qm-q)\al\om^-+s
(q^{-1}-q)^2\be\om^1\\
\om^-\ga=s\qm\ga\om^-+s(q^{-2}-1)\de\om^1 & \om^2\ga=s\ga\om^2+s(\qm-q)
\de\om^+\\
\om^-\de=s\qm\de\om^- & \om^2\de=sq^{-2}\de\om^2+s(\qm-q)\ga\om^-+s(q^
{-1}-q)^2\de\om^1
\end{array}
\]
\sk
\renewcommand{\baselinestretch}{1.2}
\noi {\sl Values and action of the generators on the q-group elements:}
\[
 \begin{array}{llll}
   \chi_1 (\al)={{s-q^2}\over {q^3-q}} &\chi_+ (\al)=0 &\chi_-(\al)=0
   &\chi_2(\al)={{s-1}\over{q-\qm}}\\
   \chi_1 (\be)=0 &\chi_+ (\be)=0 &\chi_-(\be)=-s &\chi_2(\be)=0\\
   \chi_1 (\ga)=0 &\chi_+ (\ga)=-s &\chi_-(\ga)=0 &\chi_2(\ga)=0\\
   \chi_1 (\de)={{-q^2+s(1-q^2+q^4)}\over{q^3-q}} &\chi_+ (\de)=0
    &\chi_-(\de)=0 &\chi_2(\de)={{s-q^2}\over{q^3-q}}\\
  \end{array}
\]
\[
  \begin{array}{llll}
   \chi_1*\al={{s-q^2}\over{q^3-q}}~\al &\chi_+*\al=-s\be &\chi_-*\al=0
   &\chi_2*\al={{s-1}\over{q-\qm}}~\al\\
   \chi_1*\be={{-q^2+s(1-q^2+q^4)}\over{q^3-q}}~\be &\chi_+*\be=0
   &\chi_-*\be=-s\al &\chi_2*\be={(s-q^2)\over{q^3-q}}~\be\\
   \chi_1*\ga={{s-q^2}\over{q^3-q}}~\ga &\chi_+*\ga=-s\de &\chi_-*\ga=0
   &\chi_2*\ga={{s-1}\over{q-\qm}}~\ga\\
   \chi_1*\de={{-q^2+s(1-q^2+q^4)}\over{q^3-q}}~\de &\chi_+*\de=0
   &\chi_-*\de=-s\ga &\chi_2*\de={{s-q^2}\over{q^3-q}}~\de\\
  \end{array}
\]
\sk
\noi {\sl Exterior derivatives of the basic elements of $A$:}
\[
\begin{array}{l}
 d\alpha={{s-q^2}\over{q^3-q}}\al\om^1-s\beta\om^++{{s-1}\over{q-
\qm}}\al\om^2\\
 d\beta={{-q^2+s(1-q^2+q^4)}\over{q^3-q}}\beta\om^1-s\alpha\om^-
+{{s-q^2}\over{q^3-q}}\be\om^2\\
d\gamma={{s-q^2}\over{q^3-q}}\ga\om^1-s\de\om^++{{s-1}\over{q-\qm}}\ga
\om^2\\
d\de={{-q^2+s(1-q^2+q^4)}\over{q^3-q}}\de\om^1-s\ga\om^-+{{s-q^2}\over{q
^3-q}}\de\om^2
\end{array}
\]
\noi {\sl The $\om^i$ in
terms of the exterior derivatives on $\al,\be,\ga,
\de$:}
\[
\begin{array}{l}
 \om^1={q\over{s(-q^2-q^4+s+sq^4)}} [(q^2-s)(\kappa (\al)da+\kappa
(\be)d\ga)+q^2(s-1)(\kappa (\ga)d\be + \kappa (\de) d\de)]\\
 \om^+=-{1\over s} [\kappa (\ga) d\al + \kappa (\de) d\ga]\\
\om^-=-{1\over s} [\kappa (\al) d\be + \kappa (\be) d\de]\\
\om^2={q\over{s(-q^2-q^4+s+sq^4)}}[(s-q^2-sq^2+sq^4)(\kappa (\al)
d\al+\kappa (\be) d\ga)+(q^2-s)(\kappa (\ga) d\be+\kappa (\de) d\de)]
\end{array}
\]
\vfill\eject
\renewcommand{\baselinestretch}{1.0}
\noi {\sl Lie derivative on $\om^i$:}
\[
 \begin{array}{ll}
  \chi_1*\om^1=q(q^2-1)\om^1 + (q^{-1}-q)\om^2 &\chi_+*\om^1=-q\om^-\\
  \chi_1*\om^+=-q^{-1}\om^+ &\chi_+*\om^+=-q\om^2+q^3\om^1\\
  \chi_1*\om^-=[q(q^2+1)-q^{-1}]\om^- &\chi_+*\om^-=0\\
  \chi_1*\om^2=-q(q^2-1)\om^1 - (q^{-1}-q)\om^2 &\chi_+*\om^2=q\om^-\\
 \end{array}
\]
\[
 \begin{array}{ll}
  \chi_-*\om^1=q\om^+ &\chi_2*\om^1=0\\
  \chi_-*\om^+=0 &\chi_2*\om^+=q\om^+\\
  \chi_-*\om^-=q^{-1}\om^2-q\om^1 &\chi_2*\om^-=-q^{-1}\om^-\\
  \chi_-*\om^2=-q\om^+ &\chi_2*\om^2=0\\
 \end{array}
\]

\vfill\eject

\sect{More $q$-geometry: the contraction operator and the Lie derivative}
\def\longr{\longrightarrow}
\def\invXi{{_{\mbox{\scriptsize{inv}}}}\Xi}

In section 4 we have seen that the $\chi_i$ defined by
$da = (\chi_i*a)\; \om^i$ are the quantum analogues of the tangent
vectors at the origin of the group :

\eq \chi_i \stackrel{q\rightarrow1}{\longr} \pdxi\mid_{x=0} \en

\noi and that the left-invariant vector fields $t_i$
constructed from the
$\chi_i$ are :

\eq
t_i = \chi_i* = (id \otimes \chi_i) \D
\en
\eq
t_i \stackrel{q\rightarrow1}{\longr} e^{~\mu}_i{\partial \over
{\partial x^{\mu}}}.
\en

\noi There is a one-to-one correspondence $\chi_i \leftrightarrow t_i=
\chi_i*$. In order to obtain $\chi_i$ from $\chi_i*$ we simply apply
$\epsi$ :
\eq
(\epsi \circ t_i)(a) = \epsi(id \otimes \chi_i)\D(a) =
\epsi(a_1\chi_i(a_2)) = \epsi(a_1)\chi_i(a_2) =
\chi_i(\epsi \otimes id)\D(a) =\chi_i(a)
\en
\noi[recall (\ref{prop2})].
%In other words
%the vector space $T$ generated by the basis
%$\{\chi_i\}$ is isomorphic to the vector space
%${_{\mbox{\scriptsize{inv}}}}\Xi $ generated by the left-invariant
%vectorfields $\chi_i*.$\sk
\sk
{\sl Note 1:} The vector space $T$ can also be defined intrinsically
as the space of all linear functionals from $A$ to $\Cb$ such that
$\chi(I)=0$ and $\chi(a)=0$ if $da=0$; indeed from $0=da=(\chi_i*a)\;
\om^i$ we have $\chi_i*a =0$ and applying $\epsi$ we get $\chi_i(a) =0.$
\sk

{\sl Note 2:} The vector space $T$ is a quantum Lie algebra with Lie
bracket $[\chi,\chi']$ as given in (\ref{qLie});
the vector space $\invXi$ spanned by the left-invariant
vector fields $t_i$
is also a Lie algebra
with the induced Lie bracket $[t,t'] \equiv [\chi,\chi']* .$
\sk\sk
The $*$ product of a functional with any $\tau \in
\Ga^{\otimes n}$ may be defined as
\eq
\chi * \tau \equiv (id \otimes \chi)\D_R(\tau),
\en
where the $\D_R$ acts on a generic element
$\tau = \rho^1 \otimes \rho^2 \otimes\cdots \rho^n \;\; \in \;
\Gamma^{\otimes n}$ as in (\ref{DRGaGaGa}).
%\eqa
%\D_R(\tau) = \D_R(\rho^1 \otimes \rho^2 \otimes \cdots \rho^n)
%= \D_R(\rho^1) \otimes \D_R(\rho^2)\cdots
%\otimes  \D_R(\rho^n)  \nonumber \\[1,4em]
%=\underbrace{ \rho^1_1 \otimes \rho^2_1
%\otimes\cdots \rho^n_1 }_{ \in \;\Gamma^{\otimes n}} \otimes \underbrace
%{\rho^1_2 \otimes \rho^2_2 \otimes\cdots \rho^n_2}_{ \in \;A}
%\ena
%\noi$\rho^i$ are 1-forms and $\D_R(\rho^i)
%\equiv \rho^i_1\otimes\rho^i_2
%\;\; \in\; \Gamma\otimes A.$\sk
\sk
\sk
\noi{\sl Definition}\sk
\noi We call quantum Lie derivative along the left-invariant vector field
$t = (id \otimes \chi)\D$ the operator:
\eq
\ell_t \equiv \chi *~,
\en
\noi that is
$$ \ell_t(\tau) \equiv (id \otimes \chi)\D_R (\tau) = \chi * \tau \;\;
\;\;\;\;\; \ell_t\; : \;\; \Gamma^{\otimes n} \longr \Gamma^{\otimes n}
\;.  $$
\noi For example:
\eq
\ell_{t}(a) = t(a),~~~a \in A, \label{la}
\en
\eq
\ell_{t_i}(\om^j)=(id\otimes\chi_i)
\D_R(\om^j)=\om^k\chi_i(M_k{}^j)=\C{ki}{j}\om^k, \label{lom}
\en
\noi the classical limit being evident.

The quantum Lie derivative has properties analogous to that of the
ordinary Lie derivative:

i) it is linear in $\tau$:
\eq
\ell_t(\lambda\tau+\tau')=\lambda\ell_t(\tau)+\ell_t(\tau');
\en

ii) it is linear in $t$:
\eq
\ell_{\lambda t+t'} = \lambda\ell_t + \ell_{t'},~~\la \in \Cb.
\en
By virtue of this last property we can just study $\ell_{t_i}$, where
$\{t_i\}$ is a basis of $\invXi$.\sk
{\sl Theorem}

The following relation holds:
\eq
\ell_{t_i}(\tau\otimes \tau')=\ell_{t_j}(\tau)\otimes \;f^j{}_i\ast\tau'+
     \tau\otimes \ell_{t_i}(\tau')\label{lieprod}
\en

{\sl Proof}
\begin{eqnarray*}
\ell_{t_i}(\tau\otimes \tau')&=&\\
&=&(id\otimes\chi_i)
  \D_R(\tau\otimes \tau')\\
%&=&(id\otimes\chi_i)
% \D_R(\tau)\otimes \D_R(\tau')\\
&=&(id\otimes\chi_i)(\tau_1\otimes \tau'_1\;\otimes\tau_2\tau'_2)\\
&=&(\tau_1\otimes \tau'_1)
\chi_i(\tau_2\tau'_2)
=(\tau_1\otimes \tau'_1)
 [\chi_j(\tau_2) f^j_i(\tau'_2) +
 \varepsilon(\tau_2)\chi_i(\tau'_2)]\\
&=&\tau_1\chi_j(\tau_2)
 \otimes  \tau'_1 f^j_i(\tau'_2) + \tau_1\varepsilon(\tau_2)\otimes
\tau'_1 \chi_i (\tau'_2)\\
%%% &=&(id \otimes\chi_i)\D_R(\tau)
%\otimes  (id \otimes f^j_i)\D_R(\tau') +
%%% (id \otimes \epsi)\D_R(\tau)\otimes (id \otimes\chi_i)\D_R(\tau')\\
&=&\ell_{t_j}(\tau)\otimes (id\otimes f^j_i)*
                           \tau' +\tau\otimes \ell{_{t_i}}(\tau')
\end{eqnarray*}

\noi [remember that $\chi_j(a)$ and $f{^j}_i(a)$
are $\mbox{\boldmath$C$}$ numbers].
The same argument leads to:
\eq \ell_{t_i}(a\om^j) =  \ell_{t_k}(a)
(f^k{}_i*\om^j) + a\ell_{t_i}(\om^j) \en
\eq \ell_{t_i}(\om^j a) =  \ell_{t_k}(\om^j)
(f^k{}_i*a) + \om^j\ell_{t_i}(a). \en
\noi The classical limit of (\ref{lieprod}) is easy to recover if we
remember that $\epsi * \tau = \tau$. Formulas (\ref{lieprod}),
(\ref{la}) and (\ref{lom}) uniquely define the quantum $\ell_t$, which
reduces, for \qone, to the classical Lie derivative.
\sk
{\sl Theorem:}

The Lie derivative commutes
with the exterior derivative:
\eq
\ell_{t_{i}} (d \vartheta ) = d ( \ell_{t_{i}} \vartheta ),
\;\;\;\;\;\;\;\;\;\;\;\;\;\;\vartheta : \;\;\mbox{generic form}.
\en

{\sl Proof:}
$$\ell_{t_i}(d\vartheta)=(id\otimes\chi_i)\D_R(d\vartheta)
        =(id\otimes\chi_i)(d\otimes id))\D_R(\vartheta)=$$
$$(d\otimes\chi_i)\D_R(\vartheta)=
d\vartheta_1\underbrace{\chi_i(\vartheta_2)}
      _{\in\mbox{\boldmath$C$}}=d[\vartheta_1
                       \chi_i(\vartheta_2)]=d(\ell_{t_i}\vartheta),$$

\noi where in the second equality we have used property (\ref{propd4}).
\sk

{\sl Theorem:}

The Lie derivative commutes
with the left and right actions
$\DL$ and $\DR$:
\eq
(id \otimes \ell_t)\DL(\theta)=\DL(\ell_t \theta) \label{liec}
\en
\eq
(id \otimes \ell_t)\DR(\theta)=\DR(\ell_t \theta),~~\theta \in \Ga^
{\otimes n}.
\en
\noi The proof is easy and relies on the fact that left and right
actions commute, cf. eq. (\ref{bicovariance}). In
the classical limit, eq. (\ref{liec}) becomes :
\eq
\ell_t (\ll{x} \theta) = \ll{x} ( \ell_t \theta).
\en

{\sl Note 3:} It is not difficult to prove the associativity of the
generalized $*$ product, for example that
$(\chi *\chi')*\tau = \chi *(\chi'*\tau)$. From
this property it follows that the $q$-Lie derivative is a
representation of the $q$-Lie algebra:
$$[\ell_{t} , \ell_{t'}](\tau) = \ell_{[t ,t']}(\tau),$$
where $[\ell_{t} , \ell_{t'}](\tau) \equiv [\chi , \chi']*\tau $.\sk

We now come to the construction of the contraction operator $i_t$ along
the left-invariant vector field $t$.
\sk
{\sl Definition}
\sk
The operator $i_t$ is characterized by:

\[ \begin{array}{l}
\alpha) \hspace{1cm}
        i_{t_{i}}(a)=0 \hspace{.4cm} \; a \in A \\
\beta) \hspace{1cm}
        i_{t_{i}} ( \omega^{j} )= \delta^{j}_{i} I  \\
\gamma) \hspace{1cm}
        i_{t_{i}}
        ( \omega^{i_{1} } \wedge \ldots \omega^{i_{n}} )  =
        i_{t_{j}} ( \omega^{i_{1}} ) \f{j}{i} *
        ( \omega^{i_{2} } \wedge \ldots \omega^{i_{n}} ) -
        \omega^{i_{1}} \wedge i_{t_{i}}
        ( \omega^{i_{2} } \wedge \ldots \omega^{i_{n}} ) \\
\delta) \hspace{1cm}
        i_{t_{i}} ( a \vartheta +\vartheta') =
        a i_{t_{i}} ( \vartheta ) + i_{t_i}(\vartheta')
         \;\;\;\; \vartheta ,\vartheta'\; \mbox{generic forms} \\
\varepsilon) \hspace{1cm}
        i_{\lambda^{i}t_{i}} =
        \lambda^{i} i_{ t_{i}} \hspace{.4cm}\;\;\;
        \lambda^{i} \in
        \mbox{ \boldmath$C$}
\end{array} \]

These relations uniquely define $i_t$, and its
existence is ensured by the unicity of the expansion of a
generic $n$-form on a basis of left-invariant 1-forms :
$\vartheta = a_{ i_{1} i_{2} \ldots i_{n} }
\omega^{i_{1}} \wedge \ldots \omega^{i_{n}}$.

Relation $\delta)$ expresses the $A$-linearity of $i_{t_{i}}$
(not just the $C$-linearity).\\
Relation $\gamma)$ in the commutative
limit reduces to the analog property
of the classical contraction. This relation can
be generalized by substituting
$\otimes$ to $\wedge$.\sk
\vfill\eject

{\sl Theorem:}

With the above-defined contraction operator $i_t$, the Lie
derivative can be expressed as:
\eq
\ell_{t_i}= i_{t_i}  d +d  i_{t_i}.
\en
\noi A proof of this theorem is given in Appendix B, together
with the proof of the property

\eqa
i_{t_{i}} ( \omega^{i_{1} }\wedge\ldots \omega^{i_{n}} ) &=&
  i_{t_{j}} ( \omega^{i_{1} }\wedge\ldots \omega^{i_{s}} )
\wedge
\f{j}{i} *( \omega^{i_{s+1} } \wedge  \ldots \omega^{i_{n}} )
\nonumber\\
& &+ (-1)^{s} \omega^{i_{1}} \wedge \ldots \omega^{i_{s}} \wedge
i_{t_{i}} ( \omega^{i_{s+1} } \wedge  \ldots \omega^{i_{n}} ),
\ena
where $s$ and $n$ are integers such that $ 1 \leq s < n $.

By induction on $n$ one can also prove that
\eq
(id \otimes i_t) \DL = \DL i_t
\en
\noi holds on any $n$-form. This formula $q$-generalizes the classical
commutativity of $i_t$ with the left action $\DL$, when $t$ is a left-
invariant vector field.

\vfill\eject

\sect{Softening the quantum group}

As in the classical case, we may consider the softening of the $q$-group
structure. The idea is to allow the right-hand side of the Cartan-Maurer
equations (\ref{CM}) to be nonvanishing,
i.e. to consider ``deformations" $\mu^i$ of $\om^i$ that are no longer
left-invariant. The amount of ``deformation" is measured, as in the
classical case, by a $q$-curvature two-form $R^i$:
\eq
R^i=d\mu^i + \c{jk}{i} \mu^j \we \mu^k \label{defcurv}
\en
\noi For this definition to be consistent with $d^2=0$, the following
$q$-Bianchi identities must hold:
\eq
dR^i-\c{jk}{i} R^j \we \mu^k + \c{jk}{i} \mu^j \we R^k =0;
\label{qBianchi}
\en
\noi these are easily obtained by taking the exterior derivative of
(\ref{defcurv}) and using the $q$-Jacobi identities for the $C$ structure
constants given in (\ref{cJacobi}).
\sk
The bimodule structure of the deformed $\Ga$ is assumed to be unchanged,
i.e. the commutations between elements of $A$ and elements of the
deformed $\Ga$ are unchanged. Also, the definition
(\ref{exom}) for the wedge product is still kept unaltered, so that
the commutations between the $\mu^i$ are identical to those for the $\om
^i$ given in (\ref{commom}):
\eq
\mu^i \we \mu^j = -\Z{ij}{kl} \mu^k \we \mu^l, \label{commmu}
\en
\noi with $Z$ given by (\ref{defZ}). Note that by taking the exterior
derivative of (\ref{commmu}) we can
infer the commutations of $R^i$ with $\mu^j$:
\eq
R^i \we \mu^j - \mu^i \we R^j= -\Z{ij}{kl} (R^k \we \mu^l-
\mu^k \we R^l).
\en
\noi Indeed the terms trilinear in $\mu$ that arise after using
$d\mu=R-C\mu\mu$ do cancel, since they cancel in the case $R^i=0$,
and the wedge products are unaltered.

\sk
In the constructive procedure of Section 5, we have defined the exterior
derivative to act as:
\eq
da=\lam [\tau a-a \tau] \label{da}
\en
\eq
d\theta=\lam[\tau \we \theta - (-1)^k \theta \we \tau] \label{dtheta}
\en
\noi with $\theta \in \Ga^k$. It is
now clear that eq. (\ref{dtheta}) must be modified. Indeed this
equation, with $\tau=\mu_b^{~b}$, leads to the Cartan-Maurer
equations (\ref{CartanMaurer}), since the
commutations between the $\mu^i$ just mimic those between the
left-invariant $\om^i$. Then we define the exterior differential as:
\eq
da=\lam[sa-as]\label{das}
\en
\eq
d\theta=\lam[s \we \theta - (-1)^k \theta \we s], \label{dthetas}
\en
\noi with
\eq
s=\tau + \phi.
\en
\noi It is not difficult to see that this $d$ still satisfies the usual
properties of the exterior derivative, provided
\eq
s\we s=0,
\en
\noi and that it can be extended over
the whole ``soft" exterior algebra in the
same way as in the undeformed case.
%The left and right actions
%$\DL$ and $\DR$ are still defined as in (\ref{leftco})-(\ref{rightco}),
%the $d$ derivative being now the ``soft" $d$.
\sk
{}From $s\we s=0$ we find:
\eq
\tau \we \phi + \phi \we \tau + \phi \we \phi = 0 \label{ss0}
\en
\noi since $\tau \we \tau = 0$ still holds. It is easy to compute
the curvatures, as defined in (\ref{defcurv}), in terms of $\phi$:
\eq
R^i=\lam [s \we \mu^i + \mu^i \we s] + \c{jk}{i} \mu^j \we \mu^k=
 \lam[\phi \we \mu^i + \mu^i \we \phi].
\en
\noi where the last equality is due to the fact that if $s=\tau$
the Cartan-Maurer equations hold ($ \Rightarrow R^i=0$). Similarly
we find
the curvature of $\tau$:
\eq
R(\tau)=\lam [\phi \we \tau + \tau \we \phi]=-\lam [\phi \we \phi],
\en
\noi the last equality being due to (\ref{ss0}).
\sk
A more detailed discussion on the differential calculus corresponding
to this ``soft" exterior derivative will be given in a later
publication. Here we mention that
the soft calculus allows the definition of
quantum ``diffeomorphisms":
\eq
\de_t \mu^k \equiv \ell_t \mu^k = (i_t d + di_t)\mu^k=(\nabla t)^k + i_t
R^k, \label{qdiffeomorphisms}
\en
\noi where $\nabla$ is the quantum covariant derivative whose definition
can be read off the Bianchi identities (\ref{qBianchi}) $\nabla R^k=0$.
The construction of an action, invariant under these diffeomorphisms,
proceeds as in the classical case. We refer to \cite{Cas1,Cas2} for
some preliminary applications of this formalism to the construction of
$q$-gravity and $q$-gauge theories.
\vfill\eject

\app{The derivation of two equations}

In this Appendix we derive the two equations (\ref{bic4}) and
(\ref{bic3}). Consider the exterior derivative of eq. (\ref{omb}):
\eq
d(\om^i a)=d[ (\f{i}{j} * a) \om^j]. \label{oma}
\en
The left-hand side is equal to:
\eqa
\lefteqn{
d(\om^ia)=}\nonumber\\
& &=d\om^i \we a - \om^i \we da = -\C{jk}{i} \om^j \otimes \om^k a -
\om^i \we (\chi_j * a)\om^j=\nonumber\\
& &=-\C{jk}{i}\om^j \otimes \om^k a - (\f{i}{s} * \chi_j * a)\om^s \we
\om^j=\nonumber\\
& &=-\C{jk}{i} (\f{j}{p}* \f{k}{q} * a) \om^p \otimes \om^q - (\f{i}{s}
*\chi_j * a) (\om^s \otimes \om^j - \Rhat{sj}{pq} \om^p \otimes \om^q) =
\nonumber\\
& &=[(-\C{jk}{i} \f{j}{p}  \f{k}{q} - \f{i}{p}  \chi_q + \Rhat{sj}{pq}
\f{i}{s} \chi_j)*a](\om^p\otimes \om^q) \label{A1}
\ena
The right-hand side reads:
\eqa
\lefteqn{d[(\f{i}{j} * a)\om^j] =}\nonumber\\
& &=d(\f{i}{j} * a)\om^j + (\f{i}{j} *a) d\om^j=\nonumber\\
& &=(\chi_k* \f{i}{j} * a)\om^k \we \om^j -
(\f{i}{j} * a) \C{pq}{j} \om^p
\otimes \om^q =\nonumber\\
& &=(\chi_k *\f{i}{j} * a)(\om^k \otimes \om^j -
\Rhat{kj}{pq} \om^p \otimes \om^q)-(\f{i}{j} * a) \C{pq}{j} \om^p
\otimes \om^q =\nonumber\\
& &= [(\chi_p \f{i}{q} - \Rhat{kj}{pq} \chi_k \f{i}{j} - \C{pq}{j}
\f{i}{j})*a](\om^p \otimes \om^q), \label{A2}
\ena
\noi so that we deduce the equation
\eqa
& &-\C{jk}{i} \f{j}{p}  \f{k}{q} - \f{i}{p}  \chi_q + \Rhat{sj}{pq}
\f{i}{s} \chi_j=\nonumber\\
& &=\chi_p \f{i}{q} - \Rhat{kj}{pq} \chi_k \f{i}{j} - \C{pq}{j}
\f{i}{j}. \label{A3}
\ena

We now need two lemmas.
\sk
{\sl Lemma 1}
\sk
\eq
\f{n}{l} * a \theta = (\f{n}{r} *a)(\f{r}{l} * \theta),~~~a\in A,~\theta
\in \Ga^{\otimes n}. \label{A6}
\en

{\sl Proof:}
\eqa
\lefteqn{\f{n}{l} * a\theta=}\nonumber\\
& &(id \otimes \f{n}{l})\D (a) \DR (\theta)= a_1 \theta_1 \f{n}{l}
(a_2 \theta_2)=\nonumber\\
& &a_1 \theta_1 \f{n}{r}(a_2) \f{r}{l}(\theta_2)=
a_1\f{n}{r} (a_2)\theta_1 \f{r}{l} (\theta_2)=\nonumber\\
& &(\f{n}{r} * a) \theta_1 \f{r}{l} (\theta_2) = (\f{n}{r} * a)
(\f{r}{l} *\theta). \label{A7}
\ena
\vfill\eject

{\sl Lemma 2}
\eq
\f{r}{l} * \om^j = \Rhat{rj}{kl} \om^k. \label{A8}
\en

{\sl Proof:}
\eqa
\lefteqn{\f{r}{l} * \om^j=}\nonumber\\
& &(id \otimes \f{r}{l})\DR (\om^j)= (id\otimes \f{r}{l})[\om^k \otimes
\M{k}{j}]=\nonumber\\
& &=\om^k \f{r}{l} (\M{k}{j})=\Rhat{rj}{kl}.
\ena
\sk

Consider now eq. (\ref{dhthe}) with $h=\f{n}{l}$:
\eq
d(\f{n}{l} *a)=\f{n}{l} *da. \label{A9}
\en
\noi The first member is equal to $(\chi_k*\f{n}{l}*a)\om^k$, while the
second member is:
\eqa
\f{n}{l} * da &=& \f{n}{l} * [(\chi_j * a)\om^j]=(\f{n}{r} *
\chi_j *a)(\f{r}{l} * \om^j) \nonumber\\
&=&(\f{n}{r} * \chi_j * a)( \Rhat{rj}{kl} \om^k)
\ena
\noi We have used here the two lemmas (\ref{A6}) and (\ref{A8}).
Therefore
the following equation holds:
\eq
\chi_k * \f{n}{l}= \Rhat{rj}{kl} ~\f{n}{r} * \chi_j,  \label{A10}
\en
\noi which is just eq. (\ref{bic4}). Equation (\ref{bic3}) is obtained
simply by subtracting (\ref{A10}) from eq. (\ref{A3}).
\vfill\eject

\sect{Two theorems on $i_t$ and $\ell_t$}
\sk

{\sl Theorem}\sk
The contraction operator $i_t$ satisfies:
\begin{equation}
\label{pop}
\begin{array}{lcl}
\forall \; n, \; \forall \; s : \; 1 \leq s < n, & & \\
i_{t_{i}} ( \omega^{i_{1} }
\wedge \omega^{i_{2} }\wedge\ldots \omega^{i_{n}} ) & = &
i_{t_{j}} ( \omega^{i_{1} }
\wedge \omega^{i_{2} }\wedge\ldots \omega^{i_{s}} )
\wedge \f{j}{i} *
( \omega^{i_{s+1} } \wedge  \ldots \omega^{i_{n}} ) \\
& & + (-1)^{s} \omega^{i_{1}} \wedge \ldots \omega^{i_{s}} \wedge
i_{t_{i}} ( \omega^{i_{s+1} } \wedge  \ldots \omega^{i_{n}} ).
\end{array}
\end{equation}\sk

{\sl Proof}\sk

For all $n$, when $s=1$ (\ref{pop}) is just property $\gamma)$ of the
definition of $i_{t_{i}}$ (see Section 6).
We prove the theorem by induction on $s$.
Suppose that  (\ref{pop}) be true for $s-1$; then it is true for $s$.
Indeed :
\[
\begin{array}{lcl}
i_{t_{i}} ( \omega^{i_{1} } &\wedge&  \ldots \omega^{i_{n}} ) =\\
&=&i_{t_{j}} ( \omega^{i_{1} } \wedge \ldots \omega^{i_{s-1}} )
\wedge \f{j}{i} *
( \omega^{i_s} \wedge \ldots \omega^{i_{n}} ) + \\
& & + (-1)^{s-1} \omega^{i_{1}}
\wedge \ldots \omega^{i_{s_{p}-1}} \wedge
i_{t_{i}} ( \omega^{i_{s} } \wedge  \ldots \omega^{i_{n}} ) = \\
& = &   i_{t_{j}} ( \omega^{i_{1} } \wedge \ldots \omega^{i_{s-1}} )
\wedge \f{j}{k} * \omega^{i_{s}} \wedge \f{k}{i} *
( \omega^{i_{s+1}} \wedge \ldots \omega^{i_{n}} ) + \\
& & + (-1)^{s-1} \omega^{i_{1}} \wedge \ldots \omega^{i_{s-1}} \wedge
i_{t_{j}} ( \omega^{i_{s} } ) \f{j}{i} *
( \omega^{i_{s+1} } \wedge \ldots \omega^{i_{n}} ) + \\
& & - (-1)^{s-1} \omega^{i_{1}} \wedge \ldots \omega^{i_{s-1}} \wedge
 \omega^{i_{s} }  \wedge i_{t_{i}}
( \omega^{i_{s+1} } \wedge \ldots \omega^{i_{n}} ) = \\
& = &
[ i_{t_{j}} ( \omega^{i_{1} } \wedge \ldots \omega^{i_{s-1}} )
\wedge \f{j}{k} * \omega^{i_{s}} + \\
& & + (-1)^{s-1} \omega^{i_{1}} \wedge \ldots \omega^{i_{s-1}} \wedge
i_{t_{k}} ( \omega^{i_{s}} ) ] \wedge  \f{k}{i} *
( \omega^{i_{s+1} } \wedge \ldots \omega^{i_{n}} ) + \\
& & + (-1)^{s} \omega^{i_{1}} \wedge \ldots \omega^{i_{s}} \wedge
i_{t_{i}}
( \omega^{i_{s+1} } \wedge \ldots \omega^{i_{n}} ) = \\
& = &
i_{t_{k}} ( \omega^{i_{1} } \wedge \ldots \omega^{i_{s-1}} \wedge
\omega^{i_s})
\wedge  \f{k}{i} *
( \omega^{i_{s+1} } \wedge \ldots \omega^{i_{n}} ) + \\
& & + (-1)^{s} \omega^{i_{1}} \wedge \ldots \omega^{i_{s}} \wedge
i_{t_{i}}
( \omega^{i_{s+1} } \wedge \ldots \omega^{i_{n}} );
\end{array}
\]
\noi in the last equality we have used the inductive hypothesis.

We can conclude that (\ref{pop}) is true for all $ s: \; 1 \leq s
<n$.~~~~~~~~ Q.E.D.\\[3mm]

Remembering the $A$-linearity of $i_{t_{i}}$
the subsequent generalization
is straightforward :

\[
\begin{array}{lcl}
i_{t_{a}} ( a_{ i_{1} \ldots i_{n} } \omega^{i_{1} }\wedge
\ldots \omega^{i_{n}} ) &=&
i_{t_{j}} (
a_{ i_{1} \ldots i_{n} }
\omega^{i_{1} } \wedge \ldots \omega^{i_{s}} )
\wedge
\f{j}{i} *
( \omega^{i_{s+1} } \wedge  \ldots \omega^{i_{n}} ) + \\
& & + (-1)^{s}
a_{ i_{1} \ldots i_{n} }
\omega^{i_{1}} \wedge \ldots \omega^{i_{s}} \wedge
i_{t_{i}} ( \omega^{i_{s+1} } \wedge  \ldots \omega^{i_{n}} )
\end{array}
\]
\noi with $a_{i_1\ldots i_n} \in A$.
\sk\sk\vfill\eject

\noi {\sl Theorem}
\[
\ell_{t_{i}}= i_{t_{i}} d +d i_{t_{i}}
%\hspace{1.0cm} \mbox{che scriveremo
%$\ell_{t_{a}}= i_{t_{a}} d +d i_{t_{a}} $},
\]
that is
\begin{eqnarray}
\label{54}
\forall \;
a_{ i_{1} \ldots i_{n} }
\omega^{i_{1}} \wedge \ldots \omega^{i_{n}} \; \in \Gamma^{\wedge n },
 & & \nonumber \\
\ell_{t_{i}}
( a_{ i_{1} \ldots i_{n} }
\omega^{i_{1}} \wedge \ldots \omega^{i_{n}} )
&=&
( i_{t_{i}} d +d i_{t_{i}} )
( a_{ i_{1} \ldots i_{n} }
\omega^{i_{1}} \wedge \ldots \omega^{i_{n}} ).
\end{eqnarray}

We will show this theorem by induction on the integer $n$. To do this,
we need the following:
\sk
\noi {\sl Lemma}

If $n=1$, the theorem is true, i.e.
\begin{equation}
\label{540}
\ell_{t_{i}} ( b_{k} \omega^{k} ) =
( i_{t_{i}} d +d i_{t_{i}} )
( b_{k} \omega^{k} ).
\end{equation}
First we show that:
\begin{equation}
\label{541}
\ell_{t_{i}} ( \omega^{k} ) =
( i_{t_{i}} d +d i_{t_{i}} )
\omega^{k}.
\end{equation}
We already know that $\ell_{t_i}(\om^k)=\om^j\C{ji}{k}$. The
right-hand side of (\ref{541}) yields:
\[\begin{array}{lcl}
( i_{t_{i}} d +d i_{t_{i}} )  ( \omega^{k} ) & = &
 i_{t_{i}} d\omega^{k}  +d (i_{t_{i}} \omega^{k} ) = \\
&=&-C_{nj}{}^{k}i_{t_i}(\omega^n\wedge\omega^j)=\\
& = &
- C_{nj}{}^{k} \left(
f^n{}_{i} * \omega^j  - \omega^n \delta_{i}^{j} \right) = \\
& = &
- C_{nj}{}^{k} \left[ \left( id \otimes
f^n{}_{i}  \right) \D_R (\omega^j)  -
\delta_{i}^{j} \omega^n \right] = \\
& = &
- C_{nj}{}^{k} \left[ \left( id \otimes
f^n{}_{i}  \right) \left( \omega^{\ell} \otimes M{_\ell}^{j}   \right)
- \delta_{i}^{j} \omega^n \right] = \\
& = &
- C_{nj}{}^{k} \left[
\omega^{\ell} \Rhat{nj}{\ell i}
- \delta_{i}^{j} \omega^n \right] = \\
& = &
+ C_{nj}{}^{k} \left[
\delta_{\ell}^{n} \delta_{i}^{j}
-\Rhat{nj}{\ell i}
\right]
\omega^{\ell} =  \\
& = & \C{\ell i}{k} \omega^{\ell}
\end{array}
\]
and (\ref{541}) is thus proved.

The right-hand side of (\ref{540}) gives:
\[
\begin{array}{lcl}
( i_{t_{i}} d +d i_{t_{i}} ) ( b_{k} \omega^{k} ) & = &
i_{t_{i}} \left( d b_{k} \wedge \omega^{k} + b_{k} d \omega^{k} \right) +
d \left( b_k i_{t_{i}} (\omega^k) \right) = \\
& = &
i_{t_{j}} ( d b_{k} ) f{^j}_i * \omega^{k} -
(d b_{k}) i_{t_{i}} (\omega^{k}) + \\
& &+ b_{k} i_{t_{i}} (d \omega^{k}) +
(d b_{k}) i_{t_{i}} (\omega^{k}) = \\
& = &
i_{t_{j}} ( ( \chi_n  * b_{k} ) \omega^n ) f{^j}_i * \omega^{k} +
 b_{k} i_{t_{i}} (d \omega^{k}) = \\
& = &
( \chi_n  * b_{k} ) \delta_j^n f{^j}_i * \omega^{k} +
b_{k} ( i_{t_{i}} d + d i_{t_{i}} )
\omega^{k} = \\
& = &
( \chi_n  * b_{k} ) f{^n}_i * \omega^{k} +
b_{k}  \ell_{t_{i}} (\omega^k ) = \\
& = &
\ell_{t_{n}} (b_{k} )
f{^n}_i * \omega^{k} +
b_{k}  \ell_{t_{i}} (\omega^k ) = \\
& = &
\ell_{t_{i}} (b_k \omega^k ),
\end{array}
\]
\noi and the lemma is proved. We
now finally prove the theorem.\\
Let us suppose it to be true for a ($n-1$)-form:
\begin{equation}
\ell_{t_{a}}
( a_{ i_2 \ldots i_n}
\omega^{i_{2} } \wedge  \ldots \omega^{i_{n}} ) =
( i_{t_{i}} d + d i_{t_{i}} )
( a_{ i_2 \ldots i_n}
\omega^{i_{2} } \wedge  \ldots \omega^{i_{n}} ).\nonumber
\end{equation}
Then it holds also for an $n$-form. Indeed,
the left-hand side of (\ref{54}) yields
\[
\begin{array}{l}
\ell_{t_{i}}
( a_{ i_1 \ldots i_n}
\omega^{i_{1} } \wedge  \ldots \omega^{i_{n}} ) = \\
= \ell_{t_{j}}
( a_{ i_1 \ldots i_n}
\omega^{i_{1}})  \wedge f{^j}_i *
( \omega^{i_{2}} \wedge  \ldots \omega^{i_{n}} ) +
 a_{ i_1 \ldots i_n}
\omega^{i_{1}}  \wedge \ell_{t_{i}}
( \omega^{i_{2}} \wedge  \ldots \omega^{i_{n}} )
\end{array}
\]
\noi while the right-hand side of (\ref{54}) is given by :\\
%\[
%\forall \;  \tau \in \Gamma^{\wedge}  \hspace{1.0cm}
%f{^b}_a * d \tau  =  d ( f{^b}_a *\tau )
%\]
%che deriva dalla compatibilit\`a di $d$ con $\mbox{}_{\Gamma}\phi^{\otimes}$;
%si veda fine cap.4 par.4.
\begin{eqnarray*}
\lefteqn{( i_{t_{i}} d +d i_{t_{i}} )
( a_{ i_{1} \ldots i_{n} }
\omega^{i_{1}} \wedge \ldots \omega^{i_{n}} ) = } \\
 = &&
i_{t_{i}} [ d
( a_{ i_{1} \ldots i_{n} } \omega^{i_1}) \wedge
\omega^{i_{2}} \wedge \ldots \omega^{i_{n}} -
( a_{ i_{1} \ldots i_{n} } \omega^{i_1}) \wedge
d ( \omega^{i_{2}} \wedge \ldots \omega^{i_{n}} )] + \\
& &
d [ i_{t_{j}}
( a_{ i_{1} \ldots i_{n} } \omega^{i_1})
f{^j}_i *
(\omega^{i_{2}} \wedge \ldots \omega^{i_{n}}) -
( a_{ i_{1} \ldots i_{n} } \omega^{i_1}) \wedge
i_{t_{i}}
(\omega^{i_{2}} \wedge \ldots \omega^{i_{n}}) ] = \\
=& &
 i_{t_{j}} ( d
a_{ i_{1} \ldots i_{n} } \omega^{i_1}) \wedge
f{^j}_i *
(\omega^{i_{2}} \wedge \ldots \omega^{i_{n}}) +
d ( a_{ i_{1} \ldots i_{n} } \omega^{i_1}) \wedge
i_{t_{i}}
(\omega^{i_{2}} \wedge \ldots \omega^{i_{n}}) + \\
& &
- i_{t_{j}}
(a_{ i_{1} \ldots i_{n} } \omega^{i_1})
f{^j}_i *
d (\omega^{i_{2}} \wedge \ldots \omega^{i_{n}}) +
 a_{ i_{1} \ldots i_{n} } \omega^{i_1} \wedge
i_{t_{i}}
d (\omega^{i_{2}} \wedge \ldots \omega^{i_{n}}) + \\
& &
+ d i_{t_{j}}
(a_{ i_{1} \ldots i_{n} } \omega^{i_1})
f{^j}_i *
(\omega^{i_{2}} \wedge \ldots \omega^{i_{n}}) +
i_{t_{j}} ( a_{ i_{1} \ldots i_{n} } \omega^{i_1}) \wedge
f{^j}_i *
d (\omega^{i_{2}} \wedge \ldots \omega^{i_{n}}) + \\
& &
- d(a_{ i_{1} \ldots i_{n} } \omega^{i_1})
\wedge i_{t_{i}}
(\omega^{i_{2}} \wedge \ldots \omega^{i_{n}}) +
 a_{ i_{1} \ldots i_{n} } \omega^{i_1} \wedge
d i_{t_{i}}
(\omega^{i_{2}} \wedge \ldots \omega^{i_{n}}) = \\
 =& &
[ ( i_{t_{j}} d + d i_{t_{j}} )
( a_{ i_{1} \ldots i_{n} } \omega^{i_1} ) ] \wedge
f{^j}_i   *
(\omega^{i_{2}} \wedge \ldots \omega^{i_{n}}) +\\
&&+a_{ i_{1} \ldots i_{n} } \omega^{i_1}
( i_{t_{i}} d + d i_{t_{i}} )
(\omega^{i_{2}} \wedge \ldots \omega^{i_{n}}) = \\
=& &
\ell_{t_j}
( a_{ i_{1} \ldots i_{n} } \omega^{i_1} )\wedge
f{^j}_i *
(\omega^{i_{2}} \wedge \ldots \omega^{i_{n}}) +
 a_{ i_{1} \ldots i_{n} }\omega^{i_1} \wedge
\ell_{t_i}
(\omega^{i_{2}} \wedge \ldots \omega^{i_{n}})
\end{eqnarray*}
and the theorem is proved.\\[3mm]

\vfill\eject
\sect{A collection of formulas}
\sk
We list here some useful formulas. Most have been
derived in the paper, or are particular cases of those.
\sk
\[ d\om^i = -\c{jk}{i} \om^j \we \om^k = -\C{jk}{i} \om^j \otimes \om^k
\]
\[\C{jk}{i}=(\chi_j * \chi_k)(x^i),~~~\C{jk}{i}=[\chi_j,\chi_k](x^i)=
\chi_k(\M{j}{i}) \]
\[ \f{i}{j} (\M{l}{k})=\Rhat{ik}{lj} \]
\[ \chi_i(ab)=\chi_j(a) \f{j}{i} (b) + \epsi (a) \chi_i (b) \]
\[ x^j \in A:~~\chi_i (x^j) = \de^j_i~~\epsi (x^j)=0 \]
\[ \chi_i (x^j b)=\f{j}{i} (b) \]
\[ (\chi_i * ab)=(\chi_j * a)(\f{j}{i} *b)+a(\chi_i *b) \]
\[ \chi_i * \om^j = \C{ki}{j} \om^k,~~~~\f{i}{j} * \om^k = \Rhat{ik}{lj}
\om^l \]
\[ \chi_i * \M{k}{j}=\C{li}{j} \M{k}{l},~~~~\f{i}{j} * \M{k}{l} =
\Rhat{il}{mj}  \M{k}{m} \]
\[ \chi_i * (a\theta) = (\chi_j * a)(\f{j}{i} * \theta) + a (\chi_i *
\theta)\]
\[ \chi_i * (\theta a) = (\chi_j * \theta)(\f{j}{i} * a) + \theta
 (\chi_i * a)\]
\[ \f{i}{j} * (a\theta)=(\f{i}{k} * a)(\f{k}{j} * \theta) \]
\[ \f{i}{j} * (\theta a)=(\f{i}{k} * \theta)(\f{k}{j} * a) \]
\[\ell_{t_i} (\tau \otimes \taup)= \ell_{t_j} (\tau) \otimes \f{j}{i} *
\taup + \tau \otimes \ell_{t_i} (\taup) \]
\[\M{i}{j} (a*\f{i}{k})=(\f{j}{i} * a) \M{k}{i} \]
\[ [\chi_i,\chi_j]=\C{ij}{k} \chi_k \]
\[ [[\chi_r,\chi_i],\chi_j]-\Rhat{kl}{ij} [[\chi_r,\chi_k],\chi_l]=
[\chi_r,[\chi_i,\chi_j]] \]
\[ \Dp (\chi_i)=\chi_j \otimes \f{j}{i} + \epsi \otimes \chi_i,~~~\ep
(\chi_i)=0 \]
\[\C{mn}{i} \f{m}{j} \f{n}{k} + \f{i}{j} \chi_k= \Rhat{pq}{jk} \chi_p
\f{i}{q} + \C{jk}{l} \f{i}{l} \]
\[\chi_k  \f{n}{l}=\Rhat{ij}{kl} \f{n}{i} \chi_j  \]
\vfill\eject
\[\Rhat{nm}{ij} \f{i}{p} \f{j}{q} = \f{n}{i} \f{m}{j} \Rhat{ij}{pq}\]
\[ \Dp (\f{i}{j})=\f{i}{k} \otimes \f{k}{j},~~~\ep (\f{i}{j} ) = \de^
i_j \]
\[ \kp (\f{i}{l}) \f{l}{j} = \f{i}{l} \kp (\f{l}{j})=\de^i_j \epsi,~~
  \kpm (\f{l}{i}) \f{j}{l} = \f{l}{i} \kpm (\f{j}{l})=\de^i_j
 \epsi \]
\[ \M{i}{j} \M{r}{q} \Rhat{ir}{pk} = \Rhat{jq}{ri} \M{p}{r} \M{k}{i} \]
\[ \D (\M{i}{j})=\M{i}{k} \otimes \M{k}{j},~~~~
   \epsi (\M{i}{j})=\de^i_j \]
\[\kappa (\M{i}{j})(\M{j}{l})=\M{i}{j} \kappa (\M{j}{l} ) = \de^l_i I,~~
  \kappa^{-1} (\M{j}{l})(\M{i}{j})=\M{j}{l} \kappa^{-1} (\M{i}{j} )
   = \de^l_i I \]
\[ \ell_{t} (d\tau)=d(\ell_t \tau) \]
\[ (id \otimes \ell_t) \DL (\tau) = \DL (\ell_t \tau) \]
\[ (id \otimes \ell_t) \DR (\tau) = \DR (\ell_t \tau) \]
\[ i_{t_i} (\tau \otimes \taup ) = i_{t_j} (\tau) \otimes \f{j}{i}
(\taup) + \tau \otimes i_{t_i} (\taup) \]
\[ i_{t_i} (\tau \we \taup ) = i_{t_j} (\tau) \we \f{j}{i}
(\taup) + \tau \we i_{t_i} (\taup) \]
\[ (id \otimes i_t) \DL = \DL i_t \]
\[ \ell_t (\theta) = [i_t d + d i_t] (\theta) \]
\vfill\eject

\vfill\eject
\end{document}